\documentclass[useAMS,usenatbib]{mn2e}

\usepackage{graphicx}

\title[Debris discs in the 27 Myr old open cluster IC4665]{Debris discs
  in the 27 Myr old open cluster IC4665}
\author[R. Smith, R. D. Jeffries and J. M. Oliviera]{{R. Smith\thanks{E-mail:
rs@astro.keele.ac.uk}} and R. D. Jeffries and J. M. Oliveira
  \\
Astrophysics Group, Lennard-Jones Laboratories, Keele University,
  Keele, Staffordshire, ST5 5BG} 

\begin{document}

\date{Accepted October 2010}

\pagerange{\pageref{firstpage}--\pageref{lastpage}} \pubyear{2010}

\maketitle

\label{firstpage}

\begin{abstract}
We present Spitzer IRAC and MIPS 24$\mu$m imaging of members of the
27$\pm$5Myr old open cluster IC 4665.  Models for the assembly of
terrestrial planets through planetesimal collisions and mergers
predict episodic dust debris discs at this epoch.  We determine that
42$^{+18}_{-13}$\% of the solar-type (F5-K5) cluster members have
excess emission at 24$\mu$m indicative of these debris discs, the
highest frequency of the clusters studied with Spitzer to date.  The
majority of these discs have intermediate levels of excess
($F_{24}/F_{\rm{phot}} < 2$), and no source is found to have extreme
levels of excess indicative of a recent transient event as opposed to
steady-state collisional evolution. We find no evidence of a link between
multiplicity and 24$\mu$m excess in this cluster sample.
Only the early-type star TYC424-473-1 
($T_{\rm{eff}} \sim 8420K$) has significant near-infrared excess
from 4.5$\mu$m as measured with IRAC.  Two solar-type targets have low
significance 8$\mu$m excess but no significant 24$\mu$m excess.  All
other targets show no evidence for near-infrared excess which could
indicate the presence of an optically thick primordial disc,
demonstrating that the observed 24$\mu$m excess arises from a debris
disc. \end{abstract}

\begin{keywords}
circumstellar matter -- infrared: stars.
\end{keywords}

\section{Introduction}

A key question in astronomy today is whether the Solar System's
architecture is a typical outcome of planet formation processes.
Integral to this question is whether Earth-like planets exist in other
systems.  However, the direct detection of terrestrial planets is
difficult due to the limitations of current planet detection
techniques. An alternative method for determining how the planet formation
process proceeds around other stars is through observation of the
remnants of these processes.  

Current models of planet formation propose that dusty discs around a
new star settle and km-sized planetesimals aggregate on a short
($<$1Myr) timescale \citep{weidenschillingcuzzi}.  The largest 
planetestimals undergo runaway accretion followed by oligarchic growth
resulting in tens or hundreds of 1000km-sized bodies in their own
cleared ``feeding zones'' \citep{klahr}.  These phases may take up to
a few million 
years.  Finally these planetary embryos collide and merge in a chaotic
growth phase to form a few stable terrestrial planets over 10-100Myr
(see e.g. \citealt{weidenschilling}).   
Collisions between planetesimals produce second generation dust
populations that absorb and re-radiate star light at wavelengths
$>$10$\mu$m.  In the absence of gas the lifetime of such dust is short
($<$1Myr, see e.g. \citealt{backman}) due to collisional destruction,
stellar wind drag and 
radiation pressure forces (Poynting-Robertson drag). The presence of a
mid-infrared excess thus indicates a transient source or continuous
replenishment of the dust population, a natural consequence of the
ongoing growth and development of planetary systems. 

The MIPS instrument on the \emph{Spitzer Space Telescope} allows the
detection of 24$\mu$m excess in nearby, young stars.  Excesses
detected at 24$\mu$m imply a temperature of 100-150K.  Assuming
thermal equilibrium this translates to an offset of 3-30AU around A
and early F-type stars and 0.5-3AU in F5-K5-type stars (hereafter
known as ``solar-type stars'').  These regions are precisely those in
which we may expect to find planets.  

Recent studies with Spitzer have explored the evolution of 24$\mu$m
excess in a statistical manner (see e.g. \citealt{rieke}, \citealt{su},
\citealt{siegler}, \citealt{rebull}, \citealt{carpenter}).  These
studies have tried to answer the question of why two apparently
similar stars can have very different levels of excess emission.  To
date, the clearest dependency is on age.  In A and early F-type stars
there is evidence for a peak in the upper envelope of 
excess emission at 10-20Myr before a decay in proportion with time
(see e.g. \citealt{wyattreview} and references therein).  For
solar-type stars the number of observed objects is smaller and so
correlations are harder to establish.  Based on current evidence the
decay of 24$\mu$m excess around solar-type stars appears to follow a
similar pattern to the A stars but on a timescale that is an order of
magnitude shorter (drop from 40\% to 20\% of stars with 24$\mu$m
excess occurs from 10-100Myr for solar-type stars, and from 100-500Myr
for A-type stars, see Figure 6 of \citealt{siegler}). 
In general the levels of excess emission are
also smaller around lower mass objects ($<$2 times the photosphere)
apart from around the youngest sources.  These results can be
interpreted within the framework of the evolutionary models of Kenyon
and Bromley (\citeyear{kb05,kb06}).  These suggest that 
planetesimals take longer to form at 10-30AU around A-type stars than
at $\sim$1AU around G-type stars, and thus there are copious dust
producing collisional events for 10-100Myr during planetesimal
accretion.  Around solar-type stars 100km-sized embryos may be
complete within a few Myr, and thus subsequent observed debris is
likely to have been produced in recent catastrophic collisions like
the impact that formed the Earth-Moon system (see e.g. \citealt{canup}
and references therein).  

Within this framework we would expect to find small or no excess
emission around most solar-type stars with occasional high excess
resulting from a recent massive collision in the terrestrial planet
zone (evidence for an Earth-Moon type collision has been presented
around the $\sim$12Myr old star HD172555, \citealt{lisse}).  The
majority of these collisions are expected in the age range 10-50Myr
(see e.g. \citealt{chambers}).  The distribution of excess depends on the
frequency of impacts and the timescale for removal of the resulting
dust from the system, which is in turn dependent on the dust particle
size distribution. 

In this paper we present a study of the debris disc population of the
open cluster IC 4665 (370$\pm$50pc
\citealt{mamjek}).  The cluster has reddening of $E(B-V) = 0.18 \pm
0.05$mag (\citealt{hogg&kron}; \citealt{crawford&barnes} - corresponding to
$E(V-I) = 0.23 \pm 0.06$ and $A_{\rm{V}} = 0.59 \pm 0.16$ for the intrinsic
colours of a low mass PMS star, \citealt{bessel}). Open
clusters provide a homogeneous, chemically uniform coeval population
for which debris disc incidence rates can be calculated.  IC 4665 is
one of only 5 clusters to have an age measured using the lithium
depletion boundary giving a more accurate age determination than
from H-R diagram analysis. The determined age of 27$\pm$5 Myr
\citep{manzi} makes it
a prime target for studying the debris disc population during the
period of terrestrial planet formation. 

In this paper we describe Spitzer IRAC and MIPS 24$\mu$m photometry of
confirmed members of IC 4665.  We determine those stars with excess
24$\mu$m emission based on their position in a $K_{s}-24$ vs. $V-K_s$
colour-colour diagram and on SED fitting.  We discuss how the rates
and levels of excess emission compare with other studies of solar-type
stars and how these observations constrain models of terrestrial
planet formation.

\section[\label{s:sample}]{IC 4665 Targets}

Our primary aim in this study is to determine the disc population in
the solar-type stars in IC 4665.  We base our sample on the study by
\citet{jeffries} who used fibre spectroscopy to establish cluster
membership from a sample of 452 photometric candidates.  Membership
was assessed primarily from radial velocity, giving a candidate list
of 56 stars.  This was further refined through measurements of
Ca~{\scriptsize I}  lines at 6439 and 6463{\AA} to filter out
contamination by K-giants.  Mean proper motions for 45 of the
candidate stars (taken from the NOMAD database, \citealt{nomad}) are
-0.7 $\pm$ 1.0 mas yr$^{-1}$ in RA and -6.2 $\pm$ 0.8 mas yr$^{-1}$ in
Dec.  Two stars were found to have proper motions incompatible with
cluster membership at the 3$\sigma$ level.  A final sample of 40
candidates were confirmed as low mass cluster members.  For these
targets we used the 2MASS catalogue \citep{2mass} to determine the
$K_s$ band magnitude of the targets.  Temperatures were taken from
\citet{jeffries}.  These sources are listed in Table
\ref{tab:samplejef}. 

\begin{figure}
\includegraphics[width=84mm]{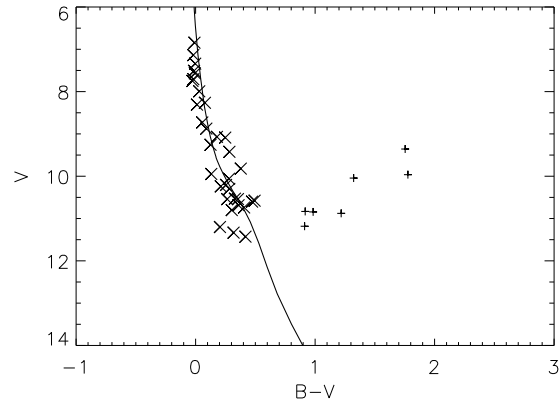}
\caption{\label{fig:brightcand} A $V$ vs $B-V$ colour-magnitude
  diagram of bright candidate members of IC 4665 from Tycho.  Here we
  show only sources with proper motions consistent with cluster
  membership.  Those sources with colours consistent with membership
  are marked by large crosses.  Overplotted is a 30Myr isochrone from
  \citet{siess} is shown with conversion following \citet{kenyon},
  adjusted for a distance of 370pc, $A_V$=0.59mag and $E(B-V)$=0.18.}
\end{figure}

To complement the low mass sample we searched the Tycho catalogue
\citep{tycho} for higher mass candidate members within a 1 degree radius
of the center of our Spitzer observations at $\rmn{RA}
17^{\rmn{h}}46^{\rmn{m}}16\fs0$, $\rmn{Dec} 5\degr41\arcmin53\arcsec$
(see section \ref{s:data} for details of the 
observed region).  A candidate list of 132 stars with 6 $<$ V $<$ 12 were 
selected.  This list was refined firstly by excluding candidates with
proper motions incompatible with the mean cluster proper motion (at
the 3$\sigma$ level as determined by \citealt{jeffries}). This reduced
the candidate high mass population to 41 stars. 
This list was further reduced by examination of the source colours in
a $B-V$ vs $V$ colour-magnitude diagram (Figure \ref{fig:brightcand}).  We
adopt a 30Myr old isochrone from \citet{siess}.  Seven stars are found
to have colours incompatible with cluster membership, and our final
list of additional bright targets comprises of 33 members.  We further
add to our list the solar-type targets P39 and P155 from the study of
IC 4665 by \citet{prosser}.  These targets were identified as possible
members based on their radial velocities and were not excluded by 
\citet{jeffries} but not observed by them
spectroscopically. Temperatures for these sources were determined from
the $B$ and $V$ magnitudes of the star $$ B - V = -3.684 \log{T_{\rm{eff}}}
+ 14.551. $$  Our final list of additional targets and their
parameters is given in Table \ref{tab:sampleoth}. 

\begin{table}
\begin{tabular}{*{6}{|c}|} 
\hline Identifier & RA & Dec & $V$ & $V-K_s$ & $T_{\rm{eff}}$, K \\
\hline 
JCO1\_427   &   266.814   &   5.94327   &   13.0690   &   2.200 &
5742 \\
JCO1\_530   &   266.885   &   5.87545   &   12.5180   &   1.655 &
6389 \\
JCO2\_145   &   266.631   &   6.09112   &   12.2720   &   1.638 &
6450 \\ 
JCO2\_213   &   266.483   &   6.04839   &   13.2180   &   2.037 &
6062 \\ 
JCO2\_220   &   266.599   &   6.04287   &   16.4200   &   4.079 &
3856 \\ 
JCO2\_373   &   266.530   &   5.93018   &   12.4560   &   2.019 &
6214 \\
JCO2\_637   &   266.662   &   6.03693   &   17.2000   &   4.936 &
3456 \\ 
JCO3\_065   &   266.379   &   6.11816   &   15.0450   &   3.377 &
4375 \\ 
JCO3\_285   &   266.149   &   5.95576   &   13.3360   &   1.721 &
6340 \\ 
JCO3\_357   &   266.112   &   5.90146   &   13.5170   &   2.476 &
5128 \\ 
JCO3\_395   &   266.354   &   5.86076   &   13.7160   &   2.280 &
5426 \\ 
JCO3\_396   &   266.242   &   5.85915   &   13.4260   &   2.086 &
5758 \\ 
JCO3\_770   &   266.112   &   5.86267   &   17.5610   &   5.105 &
3470 \\ 
JCO4\_053   &   267.021   &   5.81170   &   13.5280   &   2.663 &
5071 \\
JCO4\_226   &   266.899   &   5.64882   &   13.7810   &   2.359 &
5730 \\ 
JCO4\_337   &   266.954   &   5.54434   &   13.9130   &   3.052 &
4584 \\
JCO4\_437   &   266.735   &   5.79573   &   16.7160   &   4.364 &
3806 \\
JCO4\_459   &   266.833   &   5.78159   &   17.8930   &   4.866 &
3488  \\ 
JCO4\_591   &   266.876   &   5.68092   &   16.2360   &   3.912 &
4007 \\ 
JCO5\_179   &   266.550   &   5.69051   &   14.4070   &   2.840 &
4900 \\ 
JCO5\_280   &   266.471   &   5.60814   &   14.2430   &   2.479 &
5166 \\ 
JCO5\_282   &   266.412   &   5.60645   &   12.1460   &   1.626 &
6608 \\ 
JCO5\_296   &   266.603   &   5.59392   &   13.9400   &   2.447 &
5243 \\ 
JCO5\_472   &   266.579   &   5.77233   &   16.9100   &   4.260 &
3738 \\ 
JCO5\_515   &   266.671   &   5.73845   &   17.8510   &   5.045 &
3458 \\ 
JCO5\_521   &   266.444   &   5.73190   &   17.0620   &   4.527 &
3638 \\ 
JCO6\_088   &   266.331   &   5.79448   &   13.7160   &   2.233 &
5587 \\ 
JCO6\_095   &   266.113   &   5.78877   &   13.5110   &   2.244 &
5668 \\ 
JCO6\_111   &   266.333   &   5.77487   &   15.9930   &   3.745 &
4120 \\ 
JCO6\_240   &   266.171   &   5.69883   &   11.6580   &   1.214 &
7418 \\ 
JCO7\_021   &   266.889   &   5.52944   &   13.1260   &   2.316 &
5635 \\ 
JCO7\_079   &   266.826   &   5.50173   &   13.0880   &   1.929 &
6286 \\ 
JCO7\_088   &   266.818   &   5.49708   &   15.4110   &   3.971 &
4001 \\ 
JCO7\_670   &   266.935   &   5.36262   &   17.3490   &   5.028 &
3489 \\
JCO8\_257   &   266.603   &   5.28927   &   12.9210   &   2.109 &
5459 \\ 
JCO8\_364   &   266.648   &   5.51877   &   17.5300   &   4.847 &
3478 \\ 
JCO8\_395   &   266.394   &   5.49829   &   17.6510   &   4.710 &
3507 \\
JCO8\_550   &   266.694   &   5.37568   &   17.8720   &   5.340 &
3430 \\ 
JCO9\_120   &   266.203   &   5.48869   &   14.4050   &   2.414 &
5132 \\ 
JCO9\_281   &   266.364   &   5.40240   &   14.9560   &   3.096 &
4521 \\
\hline
\end{tabular}
\caption{\label{tab:samplejef}
Target sources in this study.  These sources, taken from
  \citet{jeffries}, are confirmed members of IC 4665.  }
\end{table}

\begin{table}
\begin{tabular}{|l|c|c|c|c|c|} 
\hline Identifier & RA & Dec & $V$ & $V-K_s$ & $T_{\rm{eff}}$, K \\
\hline 
TYC424-75-1  &   266.838  &   5.59961  &   10.739  &   0.974 &
6910 \\ 
TYC428-1339-1  &   266.759  &   5.69184  &   7.993  &   0.127 &
8740 \\ 
TYC428-1483-1  &   266.762  &   5.69863  &   10.059  &   0.579 &
7480 \\ 
TYC424-55-1  &   266.695  &   5.56493  &   8.263  &   0.402 &
8483  \\ 
TYC428-969-1  &   266.840  &   5.75952  &   9.947  &   0.772 &
8208 \\ 
TYC424-473-1  &   266.741  &   5.42569  &   8.874  &   0.410 &
8420 \\ 
TYC428-737-1  &   266.671  &   5.77427  &   7.137  &   0.074 &
9017 \\ 
TYC428-1685-1  &   266.546  &   5.65822  &   7.339  &   0.046 &
8927 \\
TYC424-1087-1  &   266.531  &   5.53021  &   6.843  &   0.025 &
8960 \\ 
TYC428-1300-1  &   267.181  &   5.70126  &   7.508 &   -0.011 &
8979 \\ 
TYC428-1571-1  &   266.488  &   5.69444  &   7.567 &   -0.020 &
8941 \\ 
TYC424-174-1  &   266.786  &   5.22533  &   10.686 &   0.899 &
7115 \\ 
TYC424-292-1  &   266.969  &   5.23321  &   11.478 &   1.464 &
6863 \\ 
TYC428-1938-1  &   266.485  &   5.81237  &   11.200 &   1.253 &
7842 \\ 
TYC428-1755-1  &   266.579  &   5.93536  &   10.601 &   1.112 &
6630 \\ 
TYC428-675-1  &   266.390  &   5.71569  &   7.707 &   0.131 &
9013 \\ 
TYC424-309-1  &   266.395  &   5.42651  &   9.087 &   0.715 &
7628 \\ 
TYC428-691-1  &   266.539  &   5.98235  &   10.531 &   0.939 &
7253 \\ 
TYC424-223-1  &   266.913  &   5.11635  &   10.240 &   0.780 &
7808 \\ 
TYC428-1910-1  &   266.322  &   5.66767  &   9.070 &   0.513 &
7968 \\ 
TYC428-215-1  &   266.652  &   6.12062  &   7.748 &  -0.041 &
9056 \\ 
TYC424-1396-1  &   266.415  &   5.19808  &   9.820 &   1.000 &
7032 \\ 
TYC424-128-1  &   266.254  &   5.52291  &   9.422 & 0.840 &
7464 \\ 
TYC424-256-1  &   267.373  &   5.24004  &   10.299  &  1.020 &
7468 \\ 
TYC428-1445-1  &   267.424  &   5.92507  &   10.800  & 0.974 &
7373 \\ 
TYC428-1211-1  &   266.945  &   6.29000  &   10.578  &  0.570 &
6542 \\ 
TYC428-847-1  &   267.043  &   6.27151  &   9.256 &   0.434 &
8239 \\ 
TYC428-840-1  &   266.838  &   6.37147  &   11.333  &   0.462 &
7303 \\ 
TYC427-1623-1  &   266.066  &   5.71430  &   8.308  &  0.266 &
8842 \\ 
TYC423-66-1  &   265.983  &   5.41341  &   10.205  &   0.570 &
7596 \\ 
TYC423-369-1  &   265.958  &   5.42050  &   8.730  &   0.260 &
8601 \\ 
TYC427-1661-1  &   266.175  &   6.23509  &   10.541  &   0.706 &
7551 \\ 
TYC428-1933-1  &   266.275  &   6.36365  &   10.541  &   1.177 &
7138 \\ 
P39  &   266.609  &   5.82858  &   12.9300  &   2.011 &
5574 \\ 
P155  &   266.934  &   5.36664  &   13.5200  &   2.294 &
5043 \\ 
\hline
\end{tabular}
\caption{\label{tab:sampleoth}
Target sources in this study. These targets from Tycho and
  Prosser (P39 and P155) are likely members of IC 4665 based on proper
  motion and colour. See text for details. }
\end{table}

\section{Spitzer Data}\label{s:data} 

Data were obtained with the Spitzer Space Telescope IRAC \citep{irac}
and MIPS \citep{mips}
 instruments under Spitzer Program P40601.  MIPS data were obtained in
scan mapping mode centered on the cluster centre of 17h46m16s
+5d41\arcmin53\arcsec.  A medium scan rate, scan leg length of
0.5$^\circ$ and 20 scan legs were used to cover an area of 50\arcmin
$\times$ 50\arcmin with an exposure of $\sim$80s at each sky position
were used in each AOR.   This AOR was performed four times to achieve
the required sensitivity.  

For IRAC the observations followed the Spitzer observing manual
instructions for a rapid shallow survey.  A 12 $\times$ 12 mapping
array was used to cover a similar area to the MIPS survey.  The AOR
used 280$\arcsec$ map steps, array orientation and a 3-point cycling
dither with medium scale factor.  High dynamic range mode with
12 second exposures gave an effective on-sky exposure of 36 seconds and
avoids saturation in the range 7 $< K <$ 13.  

The data were extracted as BCD (basic calibrated data) files from the
Spitzer archive.  These data are individually flux-calibrated array
images.  The Spitzer Science Center MOPEX package \citep{mokovoz} was
used to produce the final mosaics.  We used standard MOPEX modules.
The individual 24$\mu$m MIPS frames were flat-fielded using the
flatfield module in MOPEX.  Overlap correction was determined using
the default settings in the overlap module and the final image mosaic
consisting of all four repetitions of the AOR was constructed using
the mosaic module. Mosaics were created for each of the IRAC channels
using the overlap and mosaic modules in MOPEX under default settings. 
For details of these modules see \citet{mokovoz} or the on-line MOPEX
user's guide at
http://ssc.spitzer.caltech.edu/dataanalysistools/tools/
mopex/mopexusersguide/. 

Photometry was extracted using the APEX package from MOPEX.  For the
IRAC channels the PSF is under-sampled and thus photometry was
extracted in a circular aperture of radius 3 pixels ($\sim$3\farcs6)
with background determined in an annulus of inner radius 12 pixels and
outer radius 20 pixels ($\sim$14\farcs4 -- 24\arcsec).  Apertures were
centered on the location of each source as listed in Tables
\ref{tab:samplejef} and \ref{tab:sampleoth}.  Of the targets in Table
\ref{tab:sampleoth}, 12 fell outside the image mosaics. Photometry was
corrected for 
the array-location using correction images available online.  These
correction images were mosaiced in the same way as the data frames to
produce a correction mosaic as described in the IRAC data handbook.
Aperture corrections were taken from tabulated values in the data
handbook.  Colour corrections were applied by interpolation from
tabulated values using the effective temperatures listed in Tables
\ref{tab:samplejef} and \ref{tab:sampleoth}.  
We used the tabulated values in the IRAC data handbook to
convert the flux in Jy to magnitude.  Specifically, the zero points
used were 280.9Jy at 3.6$\mu$m, 179.7Jy at 4.5$\mu$m, 115.0Jy at
5.8$\mu$m, and 64.1Jy at 8$\mu$m.  Absolute calibration of IRAC is
stable to 1--3\% \citep{reach}.  We add this 3\% error in quadrature
to statistical background errors determined from pixel to pixel
variation in the aperture module to give a final error on the IRAC
photometry.  The final photometry is listed in Table
\ref{tab:photometry}. 

For the MIPS data the PSF is not undersampled.  We used the APEX
PRF(Point Response Function) fitting module to determine a PRF model
for the final mosaic.  As the PRF can vary for source colour we
grouped the list of targets by $V-K_s$ magnitude (from 0-6 for our
targets, grouped so $\delta(V-K_s)$=1) and used the brightest
and cleanest (no near neighbours, no bad pixels) sources in each
group as the basis for the PRF model.  The prf\_estimate module was
used to determine the PRF models for each source.  These models were
used to fit the target stars to determine the source position and
flux in the PRF photometry module in APEX.  If the source could not be
well fit (according to a $\chi^2$ analysis) with either a single or
multiple point sources (active deblend), then the PRF photometry was
determined to have  failed.  For sources which were not fit in the PRF
photometry, or where these fits were sufficiently distant ($>$1\farcs225
corresponding to 1/5 the FWHM of the average PSF for all sources)
from the input source position that the detected source was unlikely
to be our target, then aperture photometry was adopted instead.
Apertures of radius 2.6 pixels (6\farcs37) were used with annuli of
inner radius 8.16 pixels and outer radius 13.06 pixels (20\arcsec --
32\arcsec) to determine the background.  

Some of our target stars are close enough to other objects for
contamination of aperture photometry to become an issue.  To determine
a correction for contaminating sources we used the APEX module in
MOPEX to create a list of all source detections ($>3\sigma$) in the
MIPS 24$\mu$m image.  We determined the sources that were most isolated
(no other detection within $>$ 10 pixels, 24\farcs5 of the target) and
grouped them by brightness (PRF determined flux).  We then used the
aperture module to place apertures at the source and at increasing
distance from the source (in a direction away from other sources) to
determine the level of flux that would fall into an aperture at a set
distance from the source.  These profiles were found to be well fit by
a Gaussian profile with FWHM of 8\farcs7 (3.5 pixels).  Flux from the
target was $<$5\% of the flux measured in an on-source aperture at
$>$13\farcs5 and $<$1\% of the on-source flux at $>$16\farcs9.  For
each of our target sources the area within 20\arcsec was checked for
detected sources in the 24$\mu$m image, and any possible contamination
calculated by multiplying the aperture flux by the value of the
Gaussian function described above at appropriate distance.  In order
to account for the possibility that sources below the 3$\sigma$
detection threshold may be contaminating the source we also considered
the detections recorded by the APEX module in the IRAC 8$\mu$m image.
The relation between 8$\mu$m and 24$\mu$m flux was calculated by
comparing the values determined for sources detected in both images.
The median and median absolute deviation for the ratio of measured
fluxes was $F_{8\mu\rm{m}}/F_{24\mu\rm{m}} = 1.32 \pm 0.856$.  Any
sources within 20\arcsec of our target detected at 24$\mu$m had their
contamination included, the source was removed from the list of
detected sources at 8$\mu$m and then the 8$\mu$m list of detections
was checked for any additional sources within 20\arcsec.  The flux of
these sources was divided by 1.32 to get a 24$\mu$m flux and the
contamination in the aperture around our target calculated as above.
The median and median absolute deviation of the contamination for our
targets is 3\% $\pm$ 3\% of the aperture flux.  Sources with high
levels of contamination or other issues (e.g. close to edge of array)
are noted in Table \ref{tab:photometry}. PRF photometry should
mitigate against the effects of nearby contaminants through the active
deblend algorithm.  We used the above steps to determine the levels of
contamination that would be seen in aperture photometry for those
sources with PRF photometry.  This was used as a method for checking
that there were no nearby sources in the IRAC 8$\mu$m mosaic that
could have been blended with the target source in the MIPS 24$\mu$m
mosaic (as this has lower resolution).  In all cases the possible
contamination was $\leq$1.5\% the PRF photometry.  

After subtracting the contaminating flux an aperture correction of
0.729 magnitudes determined from bright targets was applied.  A colour
correction was applied for each source by interpolation from values
tabulated in the MIPS data handbook.  We used the zero point of 7.14Jy
listed in the handbook to convert the fluxes into magnitudes.
Photometric error for MIPS 24$\mu$m observations is 4\%
\citep{englebracht}.  We add this in quadrature to statistical error
returned from the aperture/APEX PRF photometry modules arising from
pixel to pixel variations to give a final error on the flux. The
final MIPS photometry is listed in Table \ref{tab:photometry}.  For 4
of our targets (JC04\_053, TYC424-174-1 TYC424-292-1 and
TYC424-1396-1) the source falls outside the MIPS 24$\mu$m mosaic and
therefore it is not listed in Table \ref{tab:photometry}.

\begin{table*}
\caption{\label{tab:photometry} Spitzer IRAC and MIPS 24$\mu$m data on
  confirmed or suspected members of IC 4665.} 
\begin{tabular}{*{9}{c}} \hline Source & [3.6] & [4.5] & [5.8] & [8.0]
  &
[24] & $K_s - [24]$ & $F_{24}/F_{\rm{phot}}$ & Comments \\  \hline 
JCO1\_427
 & 10.802 [0.033] & 10.823 [0.033] & 10.804 [0.034] & 10.806 [0.035] &
10.725 [0.100] & 0.144 & 1.093 & \\ 
JCO1\_530
 & 10.877 [0.033] & 10.860 [0.033] & 10.849 [0.034] & 10.890 [0.035] &
11.060 [0.200]\rlap{$^l$} & \llap{-}0.197\rlap{$^l$} &
0.808\rlap{$^l$} & Edge \\  
JCO2\_145
 & 10.615 [0.033] & 10.602 [0.033] & 10.608 [0.033] & 10.612 [0.034] &
10.458 [0.106] & 0.176 & 1.139 & \\ 
JCO2\_213
 & 11.147 [0.033] & 11.165 [0.033] & 11.134 [0.034] & 11.207 [0.036] &
10.802 [0.174] & 0.379 & 1.374 & XS \\
JCO2\_220
 & 12.227 [0.033] & 12.225 [0.034] & 12.194 [0.036] & 12.253 [0.051] &
12.092 [0.667]\rlap{$^l$} & 0.249\rlap{$^l$} & 0.925\rlap{$^l$} & \\ 
JCO2\_373
 & 10.363 [0.033] & 10.349 [0.033] & 10.324 [0.034] & 10.348 [0.034] &
9.882 [0.085] & 0.555 & 1.616 & XS \\
JCO2\_637
 & 11.821 [0.034] & 11.793 [0.033] & 11.786 [0.034] & 11.731 [0.042] &
11.551 [0.326]\rlap{$^l$} & 0.713\rlap{$^l$} & 1.338\rlap{$^l$} & \\ 
JCO3\_065
 & 11.815 [0.033] & 11.568 [0.033] & 11.548 [0.038] & 11.546 [0.038] &
11.234 [0.318] & 0.434 & 1.210 & \\ 
JCO3\_285
 & 11.569 [0.033] & 11.569 [0.033] & 11.569 [0.035] & 11.717 [0.041] &
11.827 [0.454]\rlap{$^l$} & \llap{-}0.212\rlap{$^l$} &
0.797\rlap{$^l$}  & Contam. \\  
JCO3\_357
 & 10.972 [0.033] & 10.954 [0.033] & 10.814 [0.034] & 10.889 [0.035] &
10.898 [0.112] & 0.143 & 1.058 & \\ 
JCO3\_395
 & 11.443 [0.033] & 11.412 [0.033] & 11.433 [0.035] & 11.425 [0.038] &
11.049 [0.122] & 0.387 & 1.354 & XS \\ 
JCO3\_396
 & 11.281 [0.033] & 11.277 [0.033] & 11.265 [0.034] & 11.261 [0.035] &
10.784 [0.104] & 0.556 & 1.616 & XS \\ 
JCO3\_770
 & 12.210 [0.034] & 12.205 [0.034] & 12.133 [0.036] & 12.148 [0.049] &
12.450 [0.935]\rlap{$^l$} & 0.006\rlap{$^l$} & 0.691\rlap{$^l$} & \\ 
JCO4\_226
 & 11.462 [0.033] & 11.377 [0.033] & 11.397 [0.035] & 11.390 [0.038] &
12.114 [0.605]\rlap{$^l$} & \llap{-}0.692\rlap{$^l$} &
0.497\rlap{$^l$}  & Contam. \\  
JCO4\_337
 & 10.780 [0.033] & 10.902 [0.033] & 10.753 [0.034] & 10.721 [0.034] &
11.817 [0.483]\rlap{$^l$} & \llap{-}0.956\rlap{$^l$} &
0.351\rlap{$^l$} & Contam. \\  
JCO4\_437
 & 12.228 [0.033] & 12.200 [0.034] & 12.160 [0.036] & 12.148 [0.049] &
11.523 [0.312] & 0.829 & 1.545 & \\ 
JCO4\_459
 & 12.740 [0.034] & 12.730 [0.034] & 12.676 [0.040] & 12.668 [0.064] &
11.911 [0.479]\rlap{$^l$} & 1.116\rlap{$^l$} & 1.948\rlap{$^l$} & \\ 
JCO4\_591
 & 12.249 [0.033] & 12.307 [0.033] & 12.256 [0.037] & 12.246 [0.045] &
11.797 [0.442]\rlap{$^l$} & 0.527 & 1.229 & \\ 
JCO5\_179
 & 11.415 [0.033] & 11.431 [0.033] & 11.424 [0.035] & 11.421 [0.036] &
11.511 [0.342]\rlap{$^l$} & 0.056\rlap{$^l$} & 0.926 \rlap{$^l$} & \\ 
JCO5\_280
 & 11.711 [0.033] & 11.756 [0.033] & 11.771 [0.035] & 11.750 [0.042] &
11.859 [0.451]\rlap{$^l$} & \llap{-}0.095\rlap{$^l$} &
0.850\rlap{$^l$} & Contam. \\  
JCO5\_282
 & 10.518 [0.033] & 10.532 [0.033] & 10.528 [0.034] & 10.532 [0.034] &
10.221 [0.104] & 0.299 & 1.276 & XS \\ 
JCO5\_296
 & 11.368 [0.033] & 11.459 [0.033] & 11.305 [0.034] & 11.336 [0.038] &
10.688 [0.152] & 0.805 & 1.952 & XS \\ 
JCO5\_472
 & 12.386 [0.034] & 12.411 [0.034] & 12.348 [0.038] & 11.957 [0.044] &
11.596 [0.343]\rlap{$^l$} & 1.054\rlap{$^l$} & 1.914\rlap{$^l$} & \\ 
JCO5\_515
 & 12.505 [0.033] & 12.465 [0.034] & 12.377 [0.038] & 12.515 [0.054] &
11.647 [0.378]\rlap{$^l$} & 1.159\rlap{$^l$} & 2.005\rlap{$^l$} & \\ 
JCO5\_521
 & 12.318 [0.034] & 12.281 [0.034] & 12.336 [0.040] & 12.293 [0.053] &
11.808 [0.446]\rlap{$^l$} & 0.727\rlap{$^l$} & 1.389\rlap{$^l$} & \\ 
JCO6\_088
 & 11.414 [0.033] & 11.423 [0.033] & 11.386 [0.034] & 11.442 [0.036] &
10.895 [0.115] & 0.588 & 1.638 & XS \\ 
JCO6\_095
 & 11.240 [0.033] & 11.239 [0.033] & 11.236 [0.034] & 11.251 [0.037] &
11.248 [0.260] & 0.019 & 0.968 & \\ 
JCO6\_111
 & 12.132 [0.033] & 12.156 [0.033] & 12.077 [0.036] & 12.048 [0.040] &
11.755 [0.433]\rlap{$^l$} & 0.493\rlap{$^l$} & 1.223\rlap{$^l$} &
Contam. \\  
JCO6\_240
 & 10.410 [0.033] & 10.403 [0.033] & 10.282 [0.033] & 10.431 [0.034] &
9.650 [0.076] & 0.794 & 2.014 & XS \\ 
JCO7\_021
 & 10.688 [0.033] & 10.686 [0.033] & 10.676 [0.034] & 10.717 [0.034] &
10.675 [0.100] & 0.135 & 1.069 & \\ 
JCO7\_079
 & 11.106 [0.033] & 11.091 [0.033] & 11.095 [0.034] & 11.109 [0.036] &
10.658 [0.146] & 0.501 & 1.537 & XS \\ 
JCO7\_088
 & 11.257 [0.033] & 11.269 [0.033] & 11.252 [0.035] & 11.230 [0.037] &
11.120 [0.222] & 0.320 & 1.005 & \\ 
JCO7\_670
 & 12.068 [0.033] & 12.053 [0.033] & 11.812 [0.036] & 12.043 [0.044] &
--  & - & -- & Edge \\ 
JCO8\_257
 & 10.691 [0.033] & 10.699 [0.033] & 10.684 [0.034] & 10.722 [0.035] &
9.857 [0.143] & 0.955 & 2.331 & XS \\ 
JCO8\_364
 & 12.465 [0.034] & 12.449 [0.033] & 12.337 [0.038] & 12.337 [0.043] &
12.383 [0.883]\rlap{$^l$} & 0.300\rlap{$^l$} & 0.920\rlap{$^l$} & \\ 
JCO8\_395
 & 12.665 [0.034] & 12.656 [0.034] & 12.567 [0.040] & 12.614 [0.065] &
14.234 [0.853]\rlap{$^l$} & \llap{-}1.293\rlap{$^l$} &
0.214\rlap{$^l$} & \\  
JCO8\_550
 & 12.297 [0.033] & 12.246 [0.034] & 12.210 [0.037] & 12.230 [0.051] &
12.183 [0.749]\rlap{$^l$} & 0.349\rlap{$^l$} & 0.935\rlap{$^l$} &  \\ 
JCO9\_120
 & 11.999 [0.033] & 12.062 [0.034] & 12.038 [0.036] & 12.012 [0.046] &
11.645 [0.379]\rlap{$^l$} & 0.346\rlap{$^l$} & 1.285\rlap{$^l$} & \\ 
JCO9\_281
 & 11.740 [0.033] & 11.808 [0.034] & 11.742 [0.035] & 11.765 [0.042] &
12.332 [0.836]\rlap{$^l$} & \llap{-}0.472 & 0.543 & \\ 
TYC424-75-1 & 9.855 [0.034] & 9.734 [0.033] & 9.643 [0.034] & 9.777
[0.034] & 9.677 [0.074] & 0.088 & 1.050 & \\ 
TYC428-1339-1 & 8.444 [0.033]\rlap{$^\ast$} & 7.989
[0.032]\rlap{$^\ast$} & 7.942 [0.032] & 7.976 [0.032] & 7.563 [0.056]
& 0.303 & 1.280 & XS \\ 
TYC428-1483-1 & 9.450 [0.033] & 9.442 [0.032] & 9.454 [0.032] & 9.480
[0.032] & 9.392 [0.067] & 0.088 & 1.051 & \\ 
TYC424-55-1 & 8.435 [0.033]\rlap{$^\ast$} & 8.085
[0.032]\rlap{$^\ast$} & 7.892 [0.032] & 7.937 [0.032] & 7.756 [0.056]
& 0.106 & 1.069 & \\  
TYC428-969-1 & 9.189 [0.033] & 9.126 [0.032] & 9.139 [0.032] & 9.174
[0.032] & 8.817 [0.059] & 0.358 & 1.347 & XS \\
TYC424-473-1 & 8.686 [0.033]\rlap{$^\ast$} & 8.456 [0.032] & 8.144
[0.032] & 7.952 [0.032] & 6.641 [0.056] & 1.823 & 5.193 & XS \\
TYC428-737-1 & 9.189 [0.033]\rlap{$^\ast$} & 8.389
[0.032]\rlap{$^\ast$} & 7.122 [0.032] & 7.166 [0.032] & 7.254
[0.056]\rlap{$^\ast$} & \llap{-}0.191\rlap{$^\ast$} &
0.813\rlap{$^\ast$} &  \\  
TYC428-1685-1 & 9.383 [0.033]\rlap{$^\ast$} & 7.716
[0.032]\rlap{$^\ast$} & 7.360 [0.032] & 7.386 [0.032] & 7.315 [0.056]
& \llap{-}0.022 & 0.952 & \\ 
TYC424-1087-1 & 8.918 [0.033]\rlap{$^\ast$} & 8.613
[0.032]\rlap{$^\ast$} & 6.901 [0.032] & 6.962 [0.032] & 7.069
[0.056]\rlap{$^\ast$} & \llap{-}0.252\rlap{$^\ast$} &
0.768\rlap{$^\ast$} & \\   
TYC428-1571-1 & 9.550 [0.033]\rlap{$^\ast$} & 7.907
[0.032]\rlap{$^\ast$} & 7.639 [0.032] & 7.703 [0.032] & 7.648 [0.056]
& \llap{-}0.061 & 0.916 & \\  
TYC428-1938-1 & 9.897 [0.033] & 9.862 [0.032] & 9.887 [0.032] & 9.923
[0.032] & 9.213 [0.066] & 0.734 & 1.904 & XS  \\
TYC428-1755-1 & 9.456 [0.033] & 9.433 [0.032] & 9.447 [0.032] & 9.474
[0.032] & 9.244 [0.065] & 0.245 & 1.219 & XS \\
TYC428-675-1 & 8.885 [0.033]\rlap{$^\ast$} & 8.026
[0.032]\rlap{$^\ast$} & 7.666 [0.032] & 7.695 [0.032] & 7.756 [0.056]
& \llap{-}0.180 & 0.821 & \\  
TYC424-309-1 & 8.669 [0.033]\rlap{$^\ast$} & 8.391 [0.032] & 8.367
[0.032] & 8.421 [0.032] & 8.351 [0.057] & 0.021 &  0.988 & \\
TYC428-691-1 & 9.573 [0.033] & 9.564 [0.032] & 9.575 [0.032] & 9.617
[0.032] & 9.583 [0.071] & 0.009 & 0.977 & \\ 
TYC428-1910-1 & 8.833 [0.033]\rlap{$^\ast$} & 8.565 [0.033] & 8.563
[0.033] & 8.608 [0.033] & 8.512 [0.058] & 0.045 & 1.010 & \\  
TYC428-215-1 & 8.391 [0.088] & 8.060 [0.032]\rlap{$^\ast$} & 7.816
[0.034] & 7.918 [0.032] & 7.918 [0.056] & \llap{-}0.129 & 0.860 & \\ 
TYC424-128-1 & 8.864 [0.033]\rlap{$^\ast$} & 8.744
[0.032]\rlap{$^\ast$} & 8.601 [0.032] & 8.650 [0.032] & 8.296 [0.057]
& 0.286 & 1.261 & XS \\ 
P39 & 10.860 [0.033] & 10.788 [0.032] & 10.787 [0.032] & 10.833
[0.032] & 10.516 [0.137] & 0.403 & 1.404 & XS \\
P155 & 11.124 [0.033] & 11.130 [0.032] & 11.116 [0.032] & 10.953
[0.032] & -- & -- & -- & Edge \\ \hline
\end{tabular} \\
$^\ast$ = Saturated photometry.  $^l$ = Signal to noise $<$ 5. XS =
Significant 24$\mu$m excess emission. Contam. = High level of
contamination in 24$\mu$m photometry. Edge = Source falls on edge of
24$\mu$m mosaic. 
\end{table*}

\section{Determination of 24$\mu$m excess}\label{s:ana_xs} 

\begin{figure*}
\includegraphics[width=125mm]{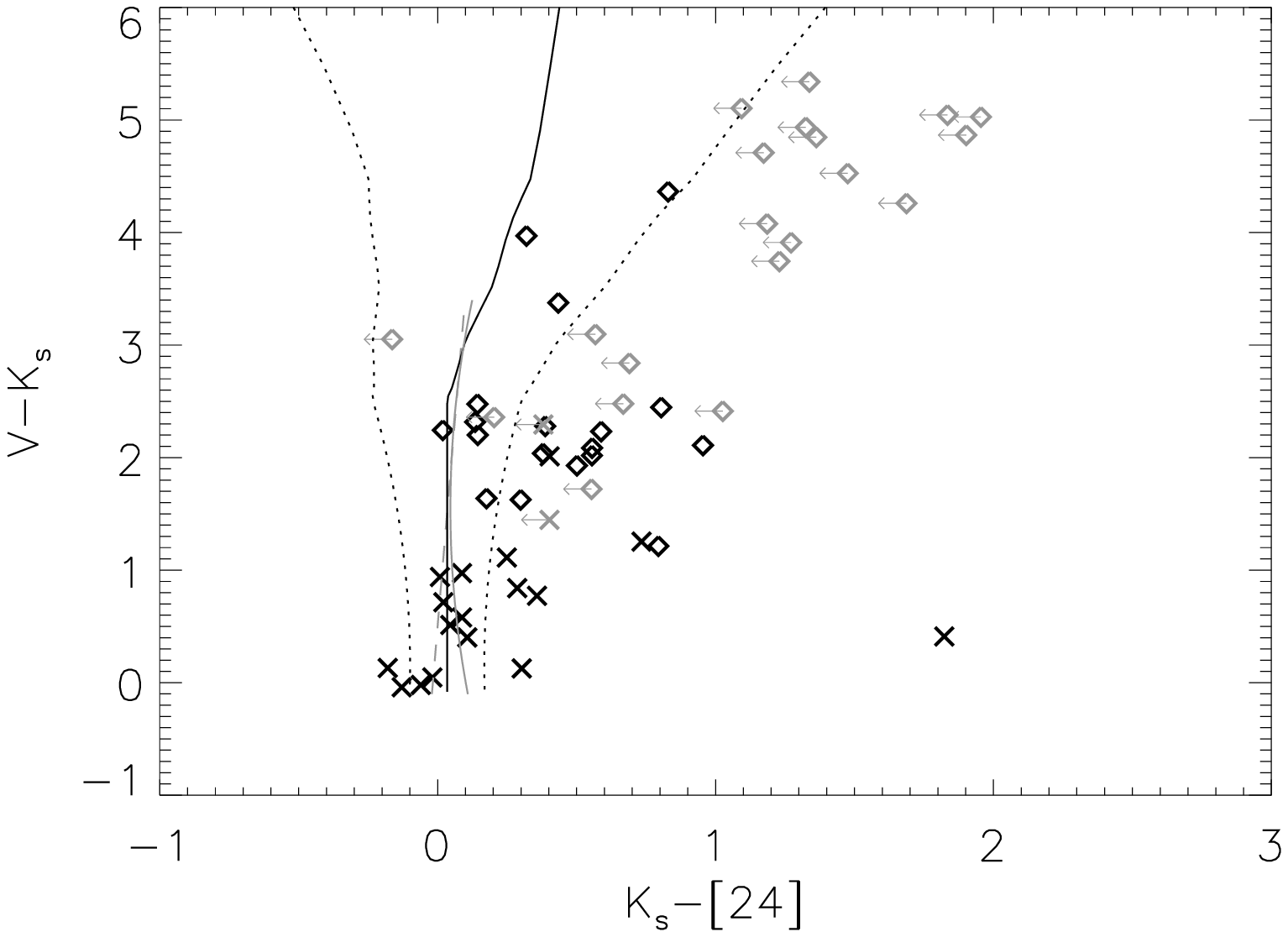}
\caption{\label{fig:colour} A colour-colour plot of the IC4665 target
  stars used to determine 24$\mu$m excess.  The solid black line is
  the expected photospheric relation from \citet{plavchan}, with
  dotted black lines showing 3$\sigma$ limits from errors on the
  24$\mu$m photometry (see text).  Also shown are the photospheric
  relations from \citet{gorlova} (dashed grey line) and
  \citet{stauffer} (solid grey line) which do not
  cover the full range of $V-K$ needed for this study. Diamond symbols
  show the targets taken from \citet{jeffries} and crosses mark the
  sources listed in Table \ref{tab:sampleoth}.  Grey symbols mark the
  5$\sigma$ upper limits for sources with low signal to noise ($<5$). }
\end{figure*}

We follow the example of recent authors \citep{rebull, stauffer} in
using the $K_s - [24]$ colour of our sources to determine which of our
targets exhibits 24$\mu$m excess emission.  This requires a well
defined model for photospheric colours.  Recently
\citet{stauffer} used Spitzer observations of the Hyades cluster to
determine an empirical relation for $K_s$ - [24] from $V - K_s$,
defined as $K_s - [24] = 0.042 - 0.053\times(V-K_s) +
0.023\times(V-K_s)^2.$ They found that this relation was very similar
to that proposed by \citet{gorlova}, but differed from the relation
given by \citet{plavchan} that included M dwarf stars.  The
Stauffer et al. relation is only valid for sources with $V - K_s <
3$.  

In Figure \ref{fig:colour} we show the $K_s-[24]$ vs $V-K_s$ colours of
our target sources.  The diamonds and crosses mark the target sources
from \citet{jeffries} as listed in Table \ref{tab:samplejef} and the
brighter targets as listed in Table \ref{tab:sampleoth} respectively.
Colours are shown in grey as upper limits if the target has a signal
to noise of less than $5\sigma$ in the 24$\mu$m photometry.
Overplotted as a solid line is the expected photospheric colour from
\citet{plavchan}.  We also show the relations from
\citet{gorlova} (dashed grey line) and from \citet{stauffer} (solid
grey line).  As our target sources cover the range $ 0 <V - K < 6 $ we
adopt the \citet{plavchan} relation as our photospheric model.
  Following the example of \citet{stauffer} the 3
  $\sigma$ errors on this relation are taken from the errors on the
  24$\mu$m photometry.  We used all significant ($>5\sigma$) 
  detections to calculate a quadratic relation between error (as given
  in Table \ref{tab:photometry}) and $V - K_s$ which was determined to
  be $ 0.045+0.004\times(V-K_s)+0.008\times(V-K_s)^2$.  The scatter around this
  fit is low, with a median difference between the fitted error and
  measured error of 2\%. 

From the figure we can identify 10 of the sources from
Jeffries et al. (JCO2\_213, JCO2\_373, JCO3\_395, JCO3\_396,
JCO5\_282, JCO5\_296, JCO6\_088, JCO6\_240, JCO7\_079 and JCO8\_257) 
and 7 of the additional targets (TYC428-1339-1, TYC428-969-1,
TYC424-473-1, TYC428-1938-1, TYC428-1755-1, TYC424-128-1, P39) that have $K_s -
[24]$ greater than the 3$\sigma$ limit on the photospheric colours.
Excluding the targets which do not fall within the MIPS 24$\mu$m mosaic
(see section \ref{s:data}) this gives a total of 16/59
sources or 27$^{+9}_{-7}$\%.  If we consider only the sources with confirmed
membership, i.e. those from Jeffries et al., we have an excess
detection frequency of 10/39 or 26$^{+11}_{-8}$\%.  Note that
  2 targets appear to have 3$\sigma$ negative detections.  These
  targets have very low $V-K_s$, in the range at which the models of
  \citet{stauffer}, \citet{gorlova}, and \citet{plavchan} most
  differ.  If we were to use the \citet{gorlova} models for the
  photospheric colours these targets would be within 3$\sigma$ of the
  expected colour, however this model does not cover the full $V-K_s$
  range of interest in this study.  Using any of the considered models
  does not change the conclusions regarding which sources have
  24$\mu$m excess.  

\subsection{Near infrared excess?}

In addition to the MIPS 24$\mu$m observations we have IRAC
photometry across 4 channels which allows us to search for evidence of
near-infrared excess.  If any stars exhibit near-infrared excess we
could be seeing the remnants of an optically thick primordial disc
rather than a true debris disc.  At the 27Myr age of this cluster such
primordial discs would be rare as they are expected to dissipate on a
timescale of a few Myr (see e.g. \citealt{wyattreview} and references
therein).  Transition discs would also be unexpected, as the
  lifetime of such discs is expected to be very short, $<$1Myr,
  \citep{skrutskie}.

\begin{figure*}
\begin{minipage}{80mm}
\includegraphics[width=80mm]{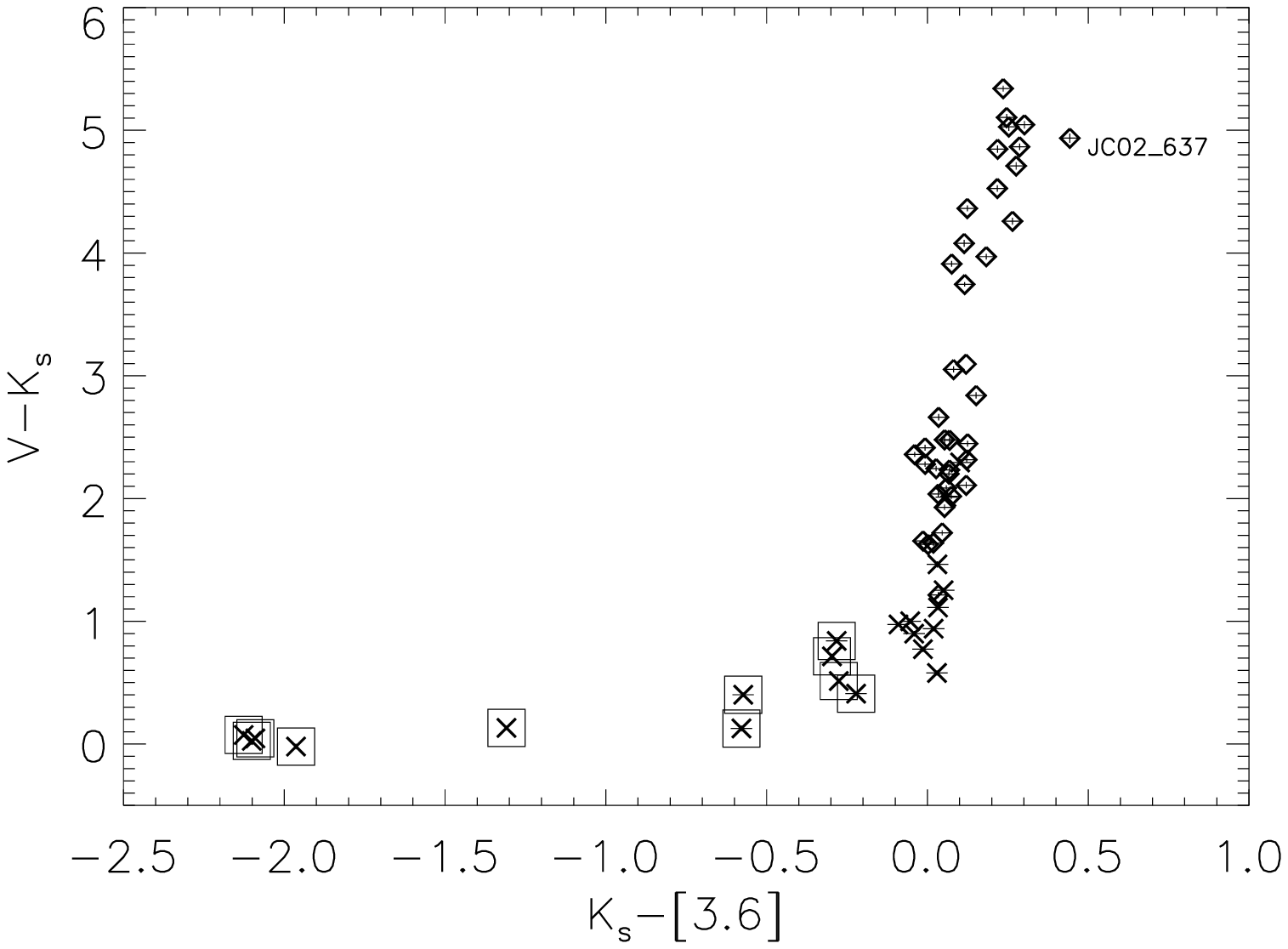}
\end{minipage}
\begin{minipage}{80mm}
\includegraphics[width=80mm]{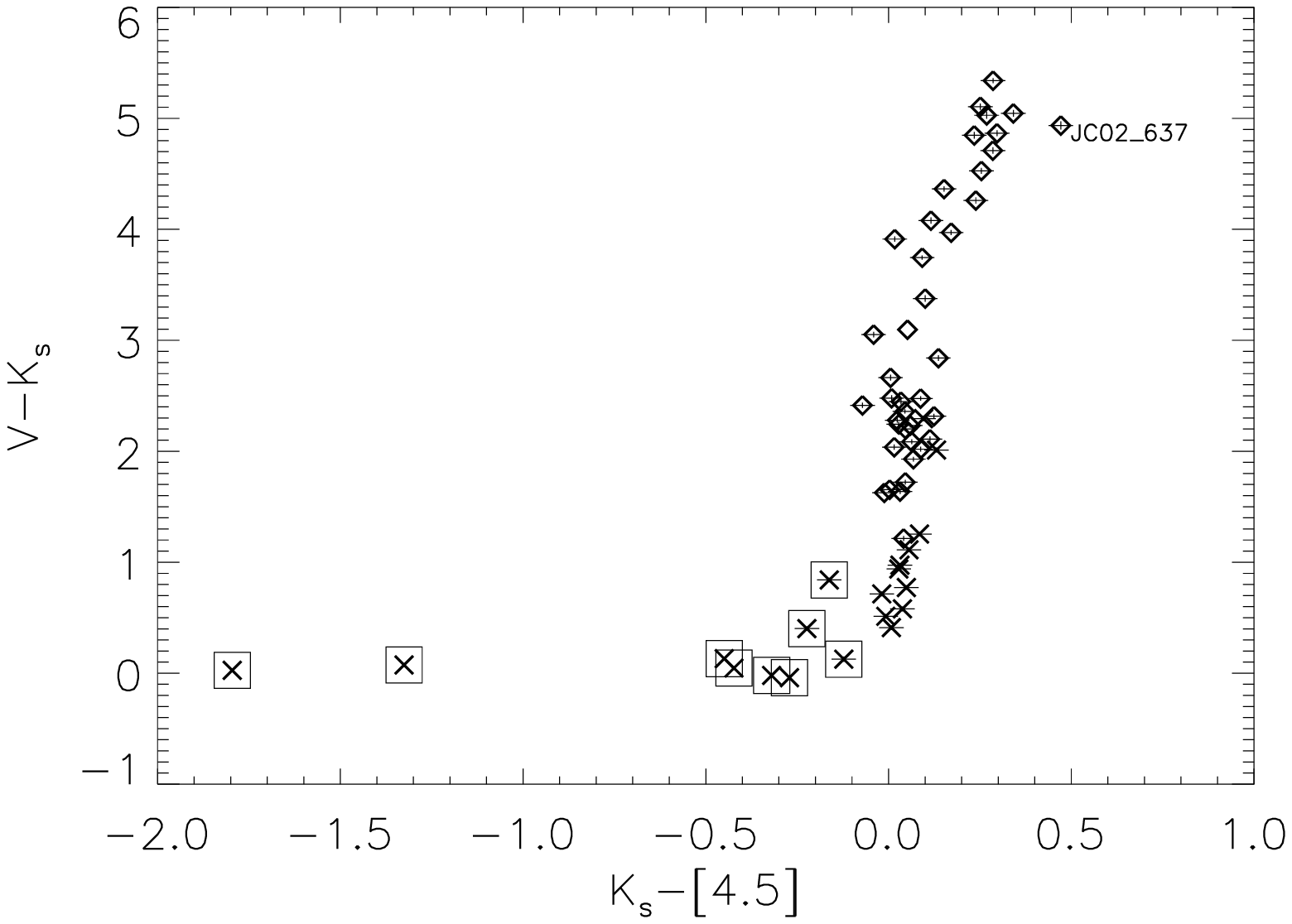}
\end{minipage} \\
\begin{minipage}{80mm}
\includegraphics[width=80mm]{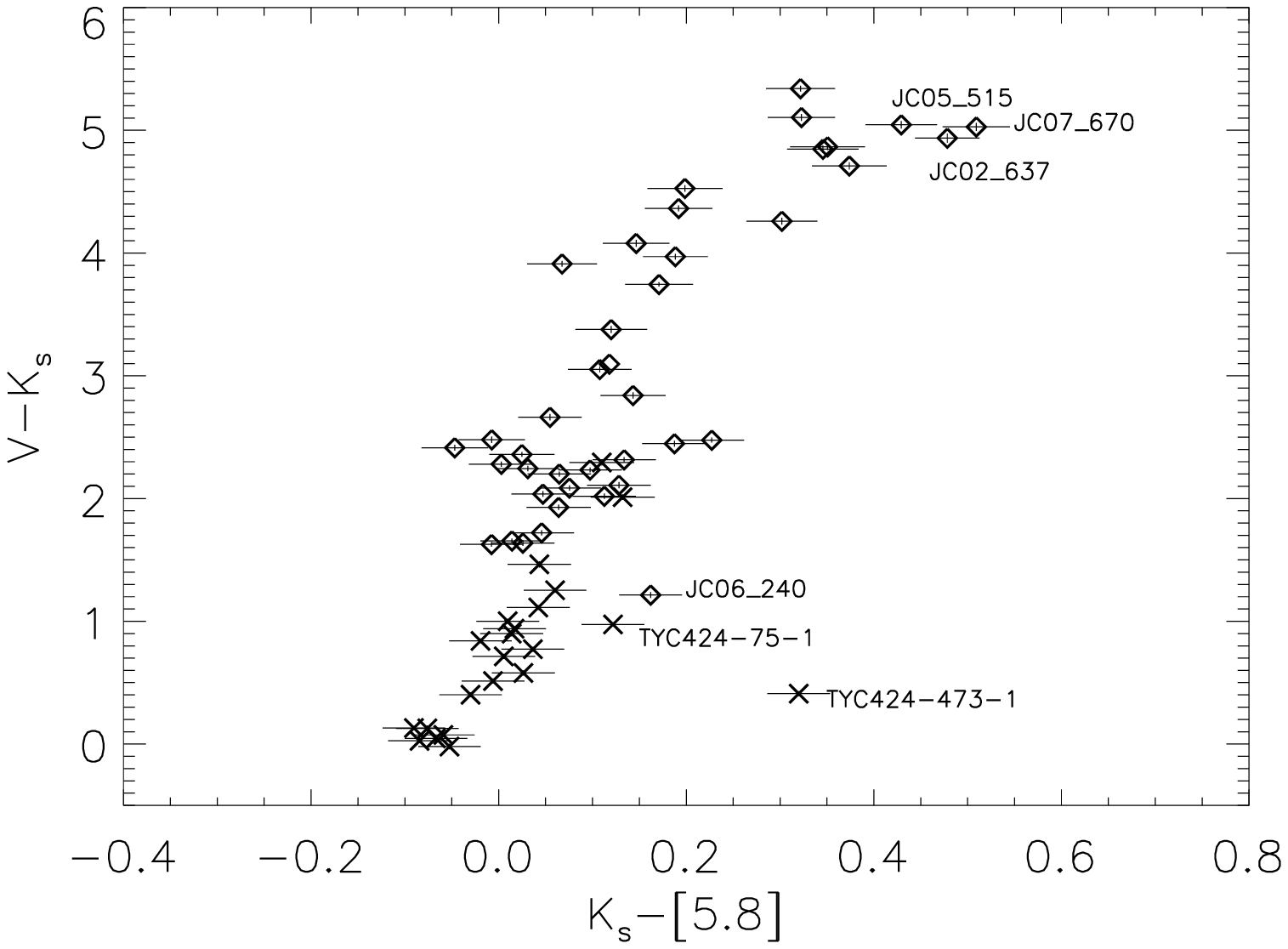}
\end{minipage}
\begin{minipage}{80mm}
\includegraphics[width=80mm]{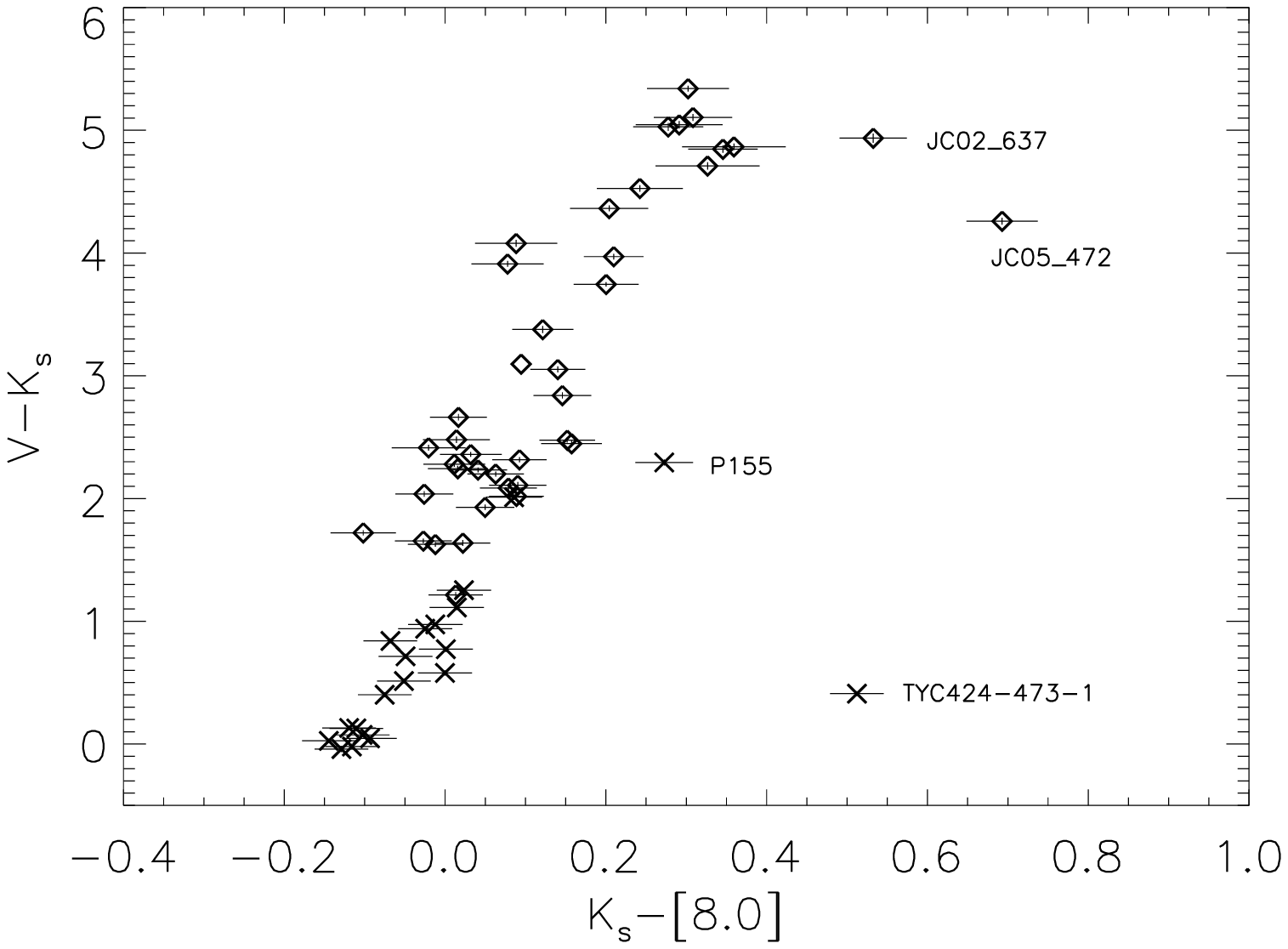}
\end{minipage} \\
\caption{\label{fig:iraccolours} A colour-colour plot of the IC4665 target
  stars for each of the IRAC channels.  Diamond symbols
  show the targets taken from \citet{jeffries} and crosses mark the
  sources listed in Table \ref{tab:sampleoth}. Sources are shown with
  1$\sigma$ errors from the IRAC photometry. Sources that appear
  saturated in the IRAC image are indicated by a surrounding square
  box.  Labelled sources have colours that may indicate excess in this
  IRAC band. See text for discussion.}
\end{figure*}

To determine if any excess emission exists in the IRAC photometry we
plot the $K_s -[3.6]$, $K_s -[4.5]$, $K_s -[5.8]$, and $K_s -[8.0]$
colours against $V - K_s$.  These plots can be seen in Figure
\ref{fig:iraccolours}.  In all panels the diamonds and crosses mark
the targets from \citet{jeffries} and those listed in Table
\ref{tab:sampleoth} respectively.  Those sources marked with square
boxes are those which are saturated in the IRAC image and therefore
have very unreliable photometry in this band (these are marked in
Table \ref{tab:photometry} with an asterisk).  Overplotted are the
errors from the IRAC photometry. Sources with colours that may
indicate excess are labelled, and we discuss these individually
below. 

\subsubsection{Poor $K_s$ photometry}
The source JC02\_637 has apparent excess in all IRAC channels. The
2MASS catalogue record for this target from which we obtain $K_s$
flags the $JHK_s$ results as poor due to issues in the PSF fit
photometry.  Furthermore if we produce an SED fit to the temperature
of the target as listed in \citet{jeffries} and scaled to the IRAC
3.6$\mu$m flux the results from the remaining channels are consistent
with photospheric emission alone.  Thus we do not believe we have
evidence for a near-infrared excess around this target.  Note that as
one of our coolest targets the 24$\mu$m MIPS detection of this source
is below a signal to noise threshold of 5, and thus we have only upper
limits on the 24$\mu$m flux. 

\subsubsection{Insignificant excess}
Two further sources, JC05\_515 and JC07\_670, with similar $V-K_s$
have possible evidence for excess at 5.8$\mu$m.  For targets in the
range $4.5 < V-K_s < 5.5 $ (excluding JC02\_637, JC05\_515 and
JC07\_670) we have an average $K_s - [5.8]$ of 0.346 $\pm$ 0.062.  The
colour of JC05\_515 is $K_s - [5.8] = 0.429 \pm 0.048$, consistent
with this range.  The colour of JC07\_670, $K_s - [5.8] = 0.509 \pm
0.048$ is closer to indicative of a significant excess. As there is no
evidence for excess in the other IRAC bands, there is no significant
evidence for a near-infrared excess which could indicate a primordial
disc remnant.  

The source JC06\_240 also has apparent 5.8$\mu$m excess.  In this case
comparison with targets of similar colour (JC06\_240 has $V-K_s =
1.214$, we look at targets with $0.7 < V-K_s < 1.7$) suggests we may
have a significant 5.8$\mu$m excess; JC06\_240 has $K_s-[5.8] = 0.162
\pm 0.040$, the average for stars of a similar colour is
0.026$\pm$0.025. However, once again there is no evidence for excess
in the other IRAC bands including at 8$\mu$m.  This source lies on the
edges of different tiles in the IRAC channel 3 imaging and pixels in
the source location have a high level of standard deviation across
different exposures, and so the error (taken from the Spitzer
uncertainty images) is underestimated.  Adopting the uncertainty from
the standard deviation between exposures instead gives $K_s - [5.8] =
0.162 \pm 0.084$.  Other sources in the $0.7 < V-K_s < 1.7$ spectral
range were checked and none were found to have a similarly high
standard deviation. 

\subsubsection{Contaminated IRAC photometry}
The target TYC424-75-1 has an apparent excess at 5.8$\mu$m, however
examination of the 5.8$\mu$m image shows that it is close to a very
bright target which contaminates the 5.8$\mu$m aperture and which is
responsible for the high $K_s - [5.8]$.  The flux from this bright
source lies just outside the apertures at 3.6 and 4.5$\mu$m, and the
lower brightness at 8.0$\mu$m means that it does not spread to
contaminate the flux of TYC424-75-1 in this band.  

\subsubsection{Significant near-infrared excess}
\begin{figure}
\includegraphics[width=84mm]{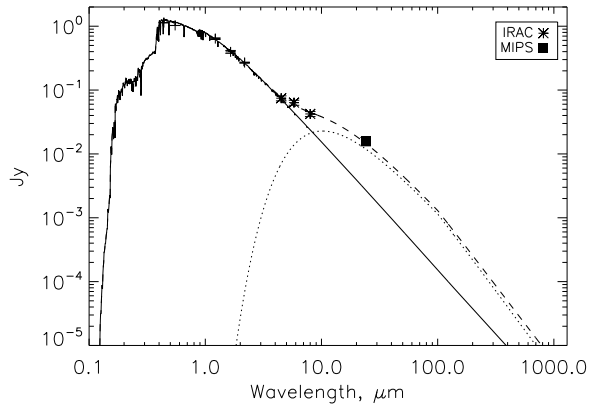}
\caption{\label{fig:TYC_xs} The SED of source TYC424-473-1.  A Kurucz
  profile of 8337K has been scaled to the 2MASS JHK photometry.
  Overplotted is a blackbody at 500K (dotted line) and the total flux
  from the photospheric model and blackbody (dashed line).  This
  profile fits the IRAC (asterisks) and MIPS (square) data we have
  obtained. }
\end{figure}

For TYC424-473-1 the difference between this and other targets of
similar colour in $V-K_s$ is highly significant.  This source is also
reddened in the $K_s - [5.8]$ plot.  This near-infrared excess could
indicate the presence of a remnant primordial disc.  We construct an SED
for this object using a Kurucz profile for a temperature of 8337K
(Figure \ref{fig:TYC_xs}).
The photospheric model is scaled to a best fit to the $JHK_s$ 2MASS
photometry using a $\chi^2$ analysis. From this plot it is clear that
the near infrared slope of the target is not consistent with
photospheric emission alone unless the target has been misidentified.
However, a single temperature blackbody added to the photospheric
emission can be shown to fit the IRAC and MIPS data.  Overplotted on
the SED is a 500K blackbody emission profile (dotted line) and the
total emission from this plus the photospheric profile (dashed line).
This total emission profile fits the Spitzer data within the errors
(note that this target is saturated in the IRAC 1 channel, and so the
[3.6] data is ignored).  The fractional luminosity of this fit
  is $f = L_{IR}/L_{\star} = 2.6\times10^{-3}$.  This is within the
  range expected for debris discs ($f<10^{-2}$, \citealt{lagrange})
  and lower than a primordial disc.  Thus we can assume the dust is
  optically thin, and thus
this temperature implies a distance of 1.7AU from the central star
assuming that the emitting material behaves like a blackbody.  If
small grains dominate the emission then the offset is likely to be
greater as such grains are inefficient emitters of radiation and thus
heat up to higher temperatures at greater distances from the star than
would be assumed from a blackbody approximation (see e.g. discussion
in Section 3 of \citealt{smithastars}). 

For two sources not yet discussed we have apparent excess at 8$\mu$m
with no excess indicated in the other IRAC bands.  JC05\_472 has a
significant excess at 8$\mu$m, with $K_s - [8.0] = 0.693 \pm 0.052$.
Other sources of similar colour ($3.8 < V-K_s < 4.8$) have a much
lower average 
$K_s - [8.0] = 0.210 \pm 0.095$.  Although this source lies close to
tile edges in the IRAC channel 4 mosaic there is no evidence of the
high standard deviation between exposures seen for JC06\_240, and thus
there is no evidence that the errors have been underestimated.  We
therefore conclude that there is evidence for an 8$\mu$m excess around
this target.  Unfortunately this source has a low level detection at
24$\mu$m ($<4\sigma$).  If we accept the low signal to noise detection
as a true reflection of the 24$\mu$m flux of the target then the flux
(164$\pm$43 mJy) is close to the expected level of flux from this
source at 24$\mu$m from an SED fit using a Kurucz profile at the
temperature taken from \citet{jeffries} and scaled to the 2MASS
$JHK_s$ photometry (expected 116mJy from the photosphere). Similarly
P155 has a large 8$\mu$m excess ($K_s - [8.0] = 0.273 \pm 0.043$
compared to an average of 0.055$\pm$ 0.052 for sources with $V-K_s =
2.3\pm0.5$).  This sources falls at the very edge of the mosaiced
24$\mu$m image which is dominated by noise and therefore we cannot
detect this target nor can we determine limits for its 24$\mu$m
flux.  There are no sources near P155 that could be contaminating the
8$\mu$m flux nor are there any indications of high variation between
different exposures including this target.  For both targets, with no
excess at shorter wavelengths and no significant detection at 24$\mu$m
the nature of the excess emission at 8$\mu$m remains currently
unresolved.  

\subsubsection{Summary}
SED plots are constructed for the other targets in our survey in the
same way, adopting a temperature as listed in Tables
\ref{tab:samplejef} and \ref{tab:sampleoth} scaled to the 2MASS JHK
photometry of the targets.  These were examined for evidence of
near-infrared excess and with the exception of the unresolved 8$\mu$m
issues for JC05\_472 and P155 (discussed above), and the excess
around TYC424-473-1, none was found.  We also
compare the targets that appear to have 24$\mu$m excess in the SED fit
to those that appear to have an excess in the $K_s-[24]$ colour plot
(Figure \ref{fig:colour}).  For all such targets the SED is consistent
with there being excess emission at 24$\mu$m.  

In summary one target, TYC424-473-1, has strong evidence of
near-infrared excess in all unsaturated observations.  For the
solar-type JC05\_472 and P155 there is some evidence of an 8$\mu$m
excess (at 5 and 4$\sigma$ significance respectively) but not of an
excess at 24$\mu$m or in other IRAC channels. 
For all other targets with uncontaminated photometry the IRAC
photometry is consistent with photospheric emission within the errors
(any excess has $<3\sigma$ significance).  The lack of IRAC
  excess indicates we do not have a population of primordial discs in
  the sample, as expected at an age of 27Myr.  These discs are also
  unlikely to be transition discs, as the lifetime of such discs is
  expected to be very short ($\sim$0.3Myr, \citealt{skrutskie}) and we
do not see a remnant primordial disc population.  Thus we consider
that the 24$\mu$m excess is indicative of a debris disc population. 

\subsection{Link with rotational velocity}

\begin{figure}
  \includegraphics[width=84mm]{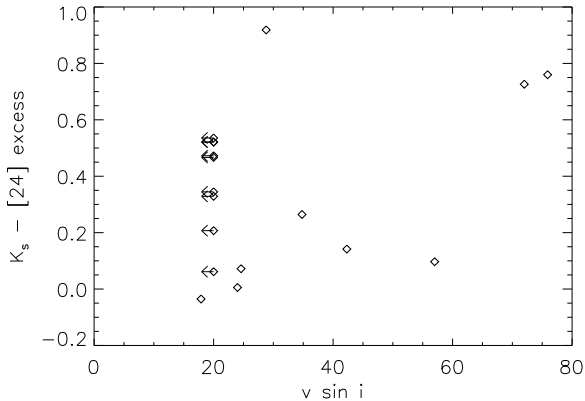}
\caption{\label{fig:rotvel} The $K_s - [24]$ excess versus projected 
  rotational velocity for the targets from \citet{jeffries}.  Here
  excess is defined by by difference from the photospheric colours
  determined by \citet{plavchan} as shown in Figure
  \ref{fig:colour}.  We find no evidence for a relationship between
  excess and rotational velocity. }
\end{figure}

Following the study by \citet{stauffer} we search for a link between 
rotational velocity and the 24$\mu$m excess.  Stellar winds from low mass
stars are thought to be powered by dynamo activity driven by
differential rotation.   Thus rotational velocity could be used as a proxy
for stellar wind.  Stellar winds are expected to remove small
particles from discs \citep{chen05, plavchan} and so we might expect a
correlation between rotational velocity and 24$\mu$m excess.  This would
not be a linear relation as the stellar wind is believed to saturate
above some rotational velocity.  Figure \ref{fig:rotvel} shows the rotational
velocity of the targets from \citet{jeffries} versus infrared excess
as measured by the difference between the photospheric relation given
by \citet{plavchan} (solid line, Figure \ref{fig:colour}) and the
measured $K_s-[24]$.  This can be compared directly with Figure 7 of
\citet{stauffer}.  We have no rotational velocity information for the
Tycho targets in our survey, and the two sources included from
\citet{prosser} have badly contaminated and thus unreliable 24$\mu$m
photometry.  

\begin{figure}
\includegraphics[width=84mm]{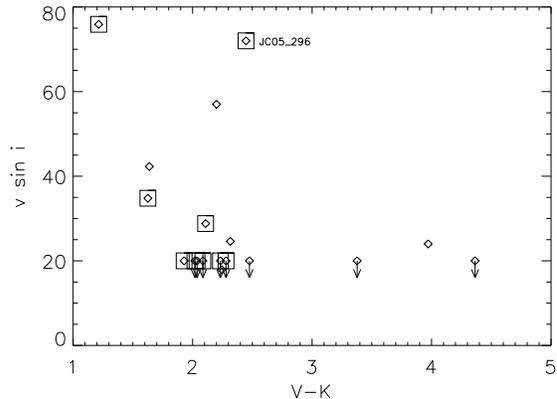}
\caption{\label{fig:rotcol} The rotational velocity for the targets
  from \citet{jeffries} shown by colour.  Squares indicate sources
  which have been judged to have 24$\mu$m excess using Figure
  \ref{fig:colour}.  One K dwarf, JC05\_296 (labelled), has a high rotational
  velocity for its spectral type and apparent 24$\mu$m excess.  This
  contradicts the expectations of the wind scouring model of
  \citet{chen05} and \citet{plavchan}. }
\end{figure}

For many of our targets the measurements of \citet{jeffries} provide
only upper limits to the rotational velocity of the targets (indicated
by arrows in Figure \ref{fig:rotvel}).  Accepting this limitation, there
is no evidence in the figure for a decrease in excess with increased
rotational velocity.  A similar null result was reported in
\citet{stauffer}, who argued that this result does not suggest
that the wind scouring model is incorrect as the sample mostly shows
the expected relation between rotational velocity and colour (lower
mass stars such as late F and G dwarfs are expected to be slow
rotators, early F type stars are rapid rotators).  Within their sample
only one rapidly rotating lower mass target was found which 
has no excess emission, and therefore the \citet{stauffer} sample
could not be used to constrain the wind scouring model.  If we examine
our targets in a similar way we find one rapidly rotating K dwarf with
an excess, as shown in Figure \ref{fig:rotcol}.  This star is
JC05\_296, a target with an excess in $K_s-[24]$ of 0.794$\pm$0.161.
As we have only one such target and suffer from 
small number statistics we cannot draw significant conclusions on the
validity of the wind scouring model based on our results.  Further
work in this area to obtain a large enough sample to test the model on
a statistical basis is required. 

\subsection{Link with multiplicity}

\begin{figure*}
\includegraphics[width=84mm]{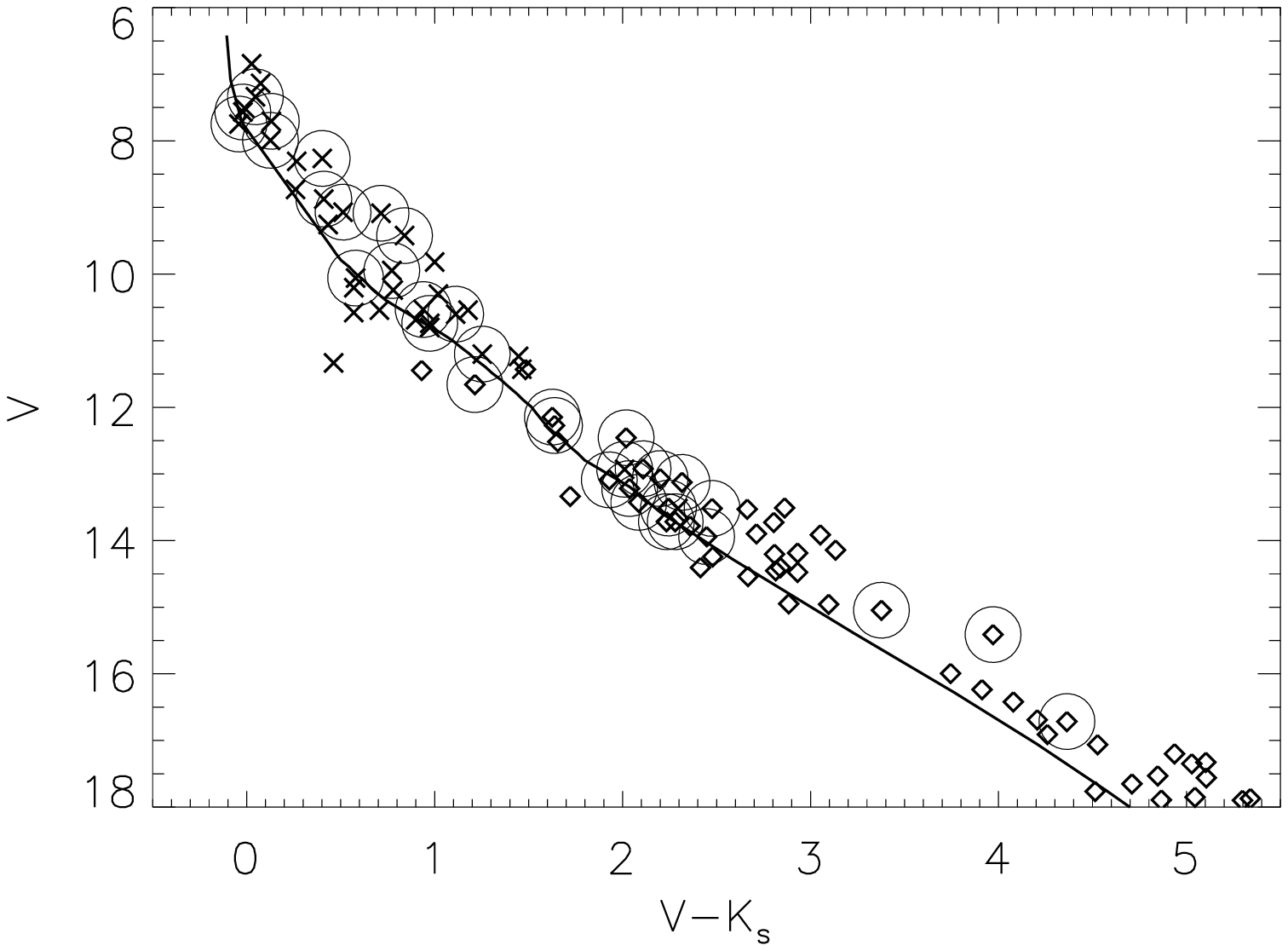}
\includegraphics[width=84mm]{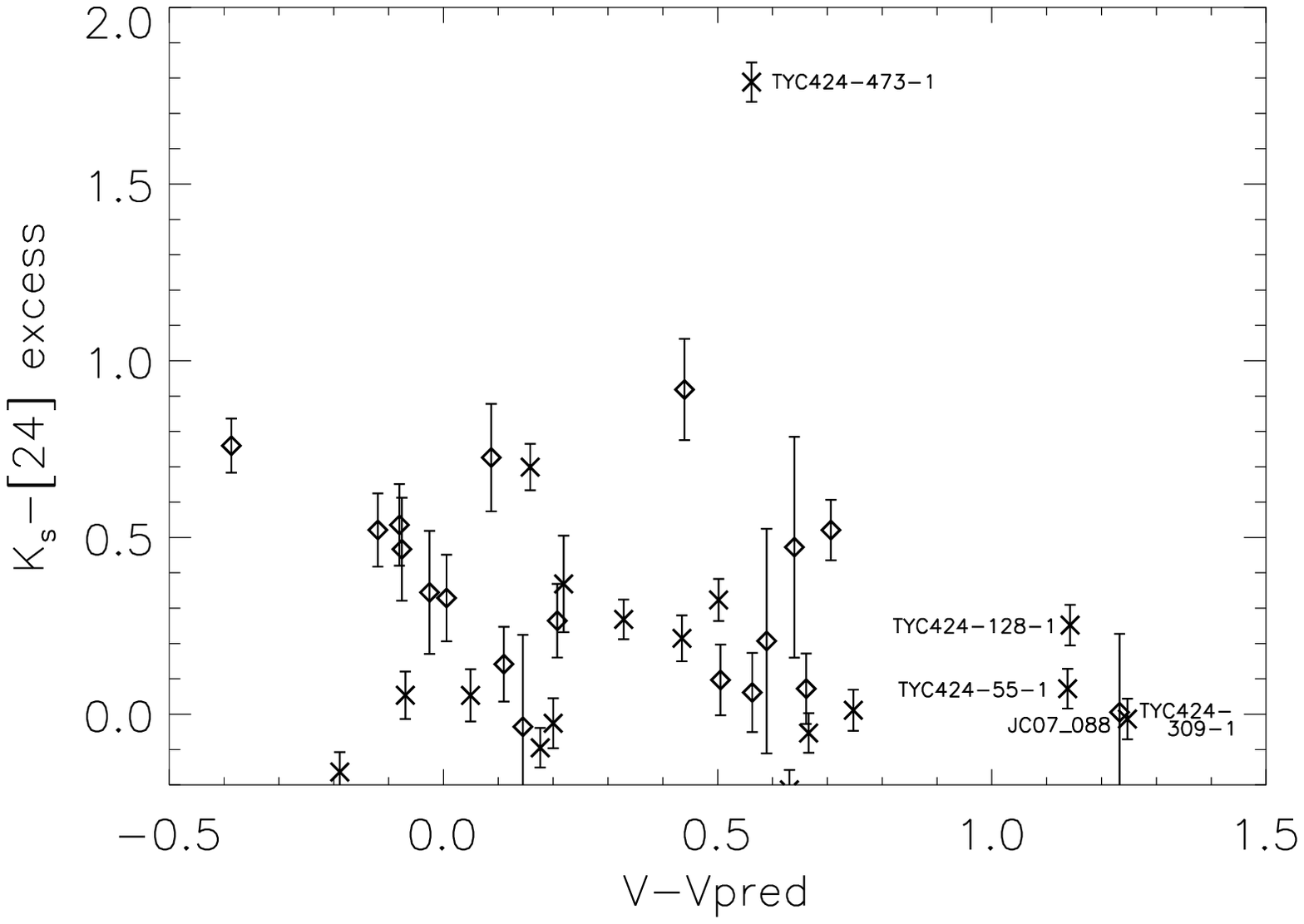}
\\
\caption{\label{fig:binarity} Dependence of excess on a proxy for
  multiplicity.  We use height above the single star isochrone (left)
  as a proxy for binarity following \citet{stauffer}.  The diamonds
  and crosses in both plots indicate targets from \citet{jeffries} and
  those listed in Table \ref{tab:sampleoth} respectively.  Stars with good
  (signal to noise $>$5) 24$\mu$m detections are indicated by
  circles. We find no evidence for a link between excess and binarity
  (right). }
\end{figure*}

\citet{cieza} presented evidence for the impact of stellar
multiplicity on the evolution of circumstellar discs.  Using the IRAC data
from several Spitzer legacy surveys they found that for projected
separations of $<$40AU systems were half as likely to retain their
primordial discs than systems with larger separations (suggesting a
disc lifetime of 0.3-0.5Myr for close binaries compared to 3-5Myr
around single stars).  Conversely
\citet{trilling} found no evidence for a link between debris disc
detection and binarity in their study of field stars (with most stars
$>$600Myr old).  This survey concentrated on detections at 70$\mu$m,
and so on discs that were further from their host stars.
\citet{plavchan} found no evidence of the trend suggested by
\citet{trilling}, and  \citet{duchene} found no significant dependence of
debris disc incidence with binarity or binary separation. 
\citet{stauffer} showed tentative evidence that in the 100Myr Blanco 1
cluster there is a link between 24$\mu$m excess and binarity (or
rather with a binarity proxy).  They combine their results with data
from the Pleiades ($\sim$ 100Myr) and NGC2547 ($\sim$ 30Myr) and find
an overall chance of 0.05\% that 
the excess around single and binary samples is drawn from the same
parent population (using a K-S test). We follow their example and use
height above a single star isochrone as a proxy for binarity and
compare this to the $K_s - [24]$ excess. The isochrone was tuned from
a fit to the Pleiades following \citet{staufferpl}. 
The results are shown in Figure \ref{fig:binarity}.  

Four targets have $V-V_{\rm{pred}}$ much higher than would indicate a binary
system (0.75 for an equal mass binary, where $V_{\rm{pred}}$ is the
predicted value of $V$ for a given value of $V - K_s$ using the single
star isochrone).  If these stars have higher
multiplicity (triple or higher order systems) then of our 35 high
signal to noise detections we would have a $11^{+9}_{-5}$\% detection
of triple or 
higher order systems.  This is compatible with previous studies (see
e.g. \citealt{abt}, \citealt{abt&levy}).  One of these systems,
TYC424-128-1, has significant excess at 24$\mu$m.  In general we find no
significant evidence for 
a dependence of excess on height above the isochrone.  Separating the
sample into those with a height above the single star isochrone of
$\ge0.3$ (in $V$) and those below this level a K-S test returns a 32\%
probability that the two populations are drawn from the same parent
distribution.  The star with the highest $K_s-[24]$ is TYC424-473-1,
the star which also has a near-infrared excess.  The value of
$V-Vpred$ found for this target (0.59) suggests that this star is a
binary and the two components are likely to be close in mass.  Using
this value to constrain our SED fitting we find a fit with a secondary
component could have a temperature of 7590K (suggesting spectral types
of A4V for the major component and A8V for the companion).  This does
not affect the fit to the excess emission presented in section 4.1.  

Amongst the sources taken from \citet{jeffries} there are two targets
with significant excess in $K_s-[24]$ that are likely binaries
(JC02\_373, $V-Vpred = 0.71$, $K_s-[24]=0.52\pm0.09$; JC08\_257,  
$V-Vpred = 0.44$, $K_s-[24]=0.92\pm0.14$).  These two sources suggest
that in this system the multiplicity of the star has no influence on
the presence of a debris disc.  In the cluster samples
  considered by \citet{stauffer} there is one large excess around a
  likely binary in the Pleiades. As we are dealing with small number
  statistics these examples do not represent a significant
  contradiction of the finding that binary and single stars have
  different debris disc populations.  However, we find no significant
  evidence for such a dependence in the IC4665 cluster.

\section{Placing IC 4665 into context}\label{s:disc} 

Following the examples of \citet{siegler} and \citet{meyer} we add the
results from the IC 4665 cluster to the growing sample of solar-type
stars studied with MIPS. These
studies and references are listed in Table \ref{tab:cluster}. Sources
included in the table are of spectral type F5-K5.  This range covers
the solar-type stars that are bright enough to have their photospheres
detected at a signal to noise of at least 3 in our MIPS 24$\mu$m
imaging.  For a 3$\sigma$ detection in the MIPS 24$\mu$m image a source is
required to have a magnitude of [24] $\leq$ 11.8.  This means we are
complete to $\sim$K5.  Concentrating on F5-K5 type stars ($1.3 \ge
V-K_s \leq 3.05$) gives us a sample of 25 stars in the MIPS field.  If we
consider only the sample restricted to this limit our excess detection
rate becomes 10/24 or 42$^{+18}_{-13}$\%.  This is somewhat higher
than other samples of a similar age, but not significantly so (see
Table \ref{tab:cluster}).  Compared to the FEPS targets presented in
\citet{meyer}, in which 
the frequency of 24$\mu$m excess was found to be 0.19$^{+0.09}_{-0.06}$
for stars aged 10-30Myr and 0.08$^{+0.06}_{-0.04}$ for stars aged
30-100Myr, IC 4665 has a reasonably large fraction of sources with
debris disc emission at 24um.  We show the frequency of 24$\mu$m excess in
the samples listed in Table \ref{tab:cluster} and the \citet{meyer}
sample in Figure \ref{fig:xsfreq}. Note that a uniform
  detection threshold ($F_{24}/F_{\rm{phot}} = 1.15$) is used in all these
  surveys with the exception of the Pleiades cluster
  \citep{sierchio}.  For this cluster we also show the frequency in
  the case that the higher threshold is used.  The detection threshold
  used for the IC4665 cluster (as given by the error calculation in
  section 4) is higher than this limit for the lower mass stars
  considered, however examination of Figure \ref{fig:colour} indicates
  that there are no objects in the region between the statistical
  errors shown and a limit of $F_{24}/F_{\rm{phot}} = 1.15$, and thus
  we consider that these results are compatible with the other cluster
  samples.  The same spectral range
(F5-K5) is considered for all clusters (and so the excess frequency
  may differ to those quoted in the papers, and in \citealt{siegler},
  where they consider a different spectral range. ) In this plot we can
see that the overall frequency of excess emission in IC 4665 is the
highest of the cluster samples listed in Table \ref{tab:cluster} but
within the errors is consistent with the frequencies seen in other
clusters of similar ages.

\begin{table*}
\caption{\label{tab:cluster} Frequency of 24$\mu$m excess from cluster
  data (after \citealt{siegler})}
\begin{tabular}{ccccl} \hline Name & Age, Myr & Number of sources &
  Excess frequency & Reference \\ \hline 
Sco Cen & 16$\pm$2 & 15 &
  0.33$^{+0.23}_{-0.14}$ & \citet{chen05} \\
\textbf{IC 4665} & \textbf{27$\pm$5} & \textbf{24} &
  \textbf{0.42$^{+0.18}_{-0.13}$} & \textbf{This paper} \\    
NGC 2547 & 30$\pm$5 & 39 & 0.33$^{+0.13}_{-0.09}$ & \citet{gorlova07}
  \\
IC 2391 & 50$\pm$5 & 13 & 0.23$^{+0.22}_{-0.13}$ & \citet{siegler} \\ 
Blanco 1 & 100$\pm$20 & 26 & 0.19$^{+0.13}_{-0.08}$ & \citet{stauffer}
  \\ 
Pleiades & 115$\pm$20 & 71 & 0.32$^{+0.08}_{-0.07}$\rlap{$^a$} &
  \citet{sierchio} \\
Hyades & 625$\pm$50 & 63 & 0.00$\pm$0.04 & \citet{cieza08} \\
Field stars & $\sim$4000\rlap{$^b$} & 69 & 0.03$^{+0.04}_{-0.02}$ &
\citet{bryden} \\ 
\hline 
\end{tabular} \\
\begin{flushleft}
 $^a$ Uses an excess detection threshold of 1.10. Majority of other
 surveys use a threshold of 1.15 (if adopting this higher threshold
 excess frequency becomes 0.21$^{+0.07}_{-0.05}$).

 $^b$ Median age of the sample 
\end{flushleft}
\end{table*}

\begin{figure}
\includegraphics[width=84mm]{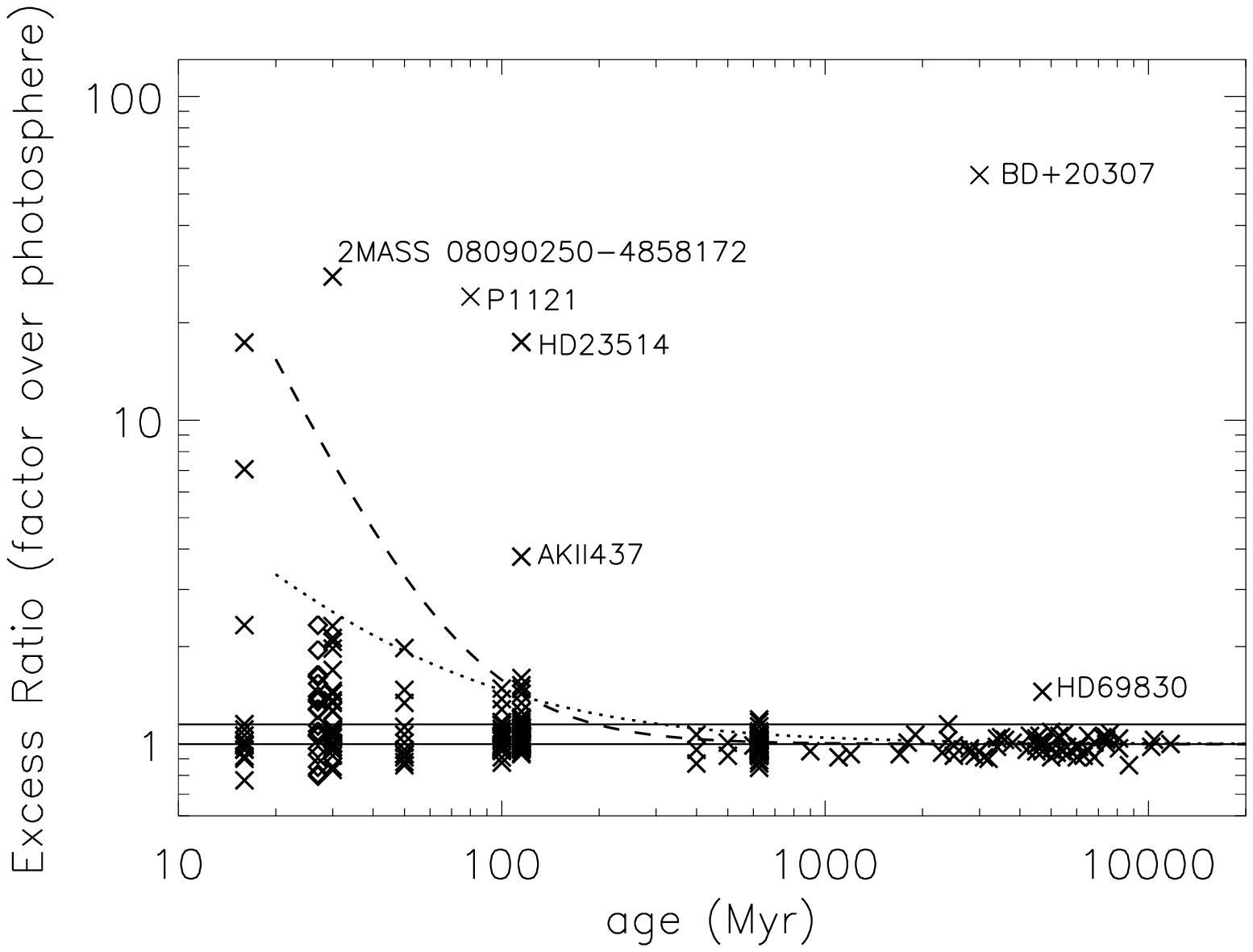}
\caption{\label{fig:fgk_xs} The 24$\mu$m excess emission around stars
  in the studies listed in Table \ref{tab:cluster}.  The sources from
  this paper are marked with diamonds.  Sources with high excess
  compared to typical ranges for their age are named.  In addition to
  the cluster data listed in Table \ref{tab:cluster} we show BD+20307
  (\citealt{song}, \textbf{age several Gyr from
  \citealt{zuckerman08}}) and P1121 in M47 \citep{gorlova}. Overplotted are 
  an inverse time dependence (dotted line) and an inverse time-squared
  dependence (dashed line).}
\end{figure}

\begin{figure}
  \includegraphics[width=84mm]{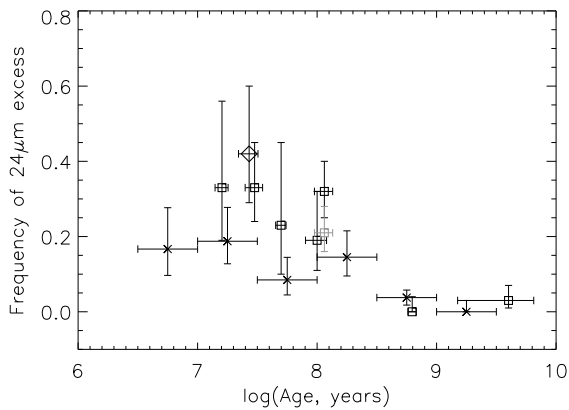}
\caption{\label{fig:xsfreq} The frequency of 24$\mu$m excess emission
  in the samples listed in Table \ref{tab:cluster} (squares) and in
  the \citet{meyer} sample (crosses).  The IC 4665 cluster data
  presented in this paper is marked with a diamond.  The grey square
  marks the excess fraction seen in the Pleiades if we adopt an excess
  detection level of $F/F\ast = 1.15$ as is typical for the other
  samples in the plot (see Table \ref{tab:cluster}.}
\end{figure}

In Figure \ref{fig:fgk_xs} we show the level of 24$\mu$m excess
(expressed as measured flux over that expected from the photosphere)
seen in studies of different clusters.  Overplotted are an
inverse time and an inverse time-squared dependence.  As first
suggested by \citet{siegler}, the inverse time-squared dependence
seems to offer the best fit to the upper envelope of excess for the
Upper Scorpious and Lower 
Centaurus Crux members of the Scorpius-Centaurus association (data
from \citealt{chen05}), but the inverse time dependence is a better
fit to the rest of the data.  An inverse time dependence has also been
shown to fit A star debris disc statistics \citep{rieke}.  This
agrees with predictions for steady-state evolution of debris discs, in
which the excess emission arising from the debris discs falls
inversely in
proportion to time. This $time^{-1}$ fall-off arises once the debris
reaches a collisional equilibrium in which the largest bodies in the
population are colliding and producing smaller material which is
eventually collisionally processed into dust small enough to be
removed from the system (\citealt{dd03, wyattsmith06}).  

Compared to the 30Myr old cluster NGC 2547 the excess levels seen in
IC 4665 show a similar spread apart from the presence of one high
excess source in NGC 2547, 2MASS 08090250-4858172 (labelled ID 8 in
\citealt{gorlova}; this source has excess emission from 3$\mu$m and is
possibly a primordial optically thick disc).  Our
results are consistent with the inverse time dependence.  We find no
sources that exceed this envelope and could be considered to have dust
resulting from a transient event as has been postulated for HD69830
and BD+20307 \citep{wyattsmith06}.  Extreme levels of excess could be
interpreted as evidence for a recent massive collision such as those
expected between proto-planets in the final stages of terrestrial
planet accretion \citep{weidenschilling}.  Simulations by
  \citet{kb05} have shown that at an age of 27Myr a collisional
  cascade in a minimum-mass solar nebula around a solar-type star at
  0.4--2AU would be above our detection threshold, and so could be a
  valid model for the emission we detect.  Similarly around higher
  mass stars the emission could arise from a collisional cascade at
  3--20AU or 30--150AU.  With a single photometric detection for most
  of these targets we cannot constrain either the location or the
  mass of the dust, however these models show that a collisional
  cascade produced by catastrophic collisions in a disc is a possible
  model for the emission seen here.

\begin{figure}
  \includegraphics[width=84mm]{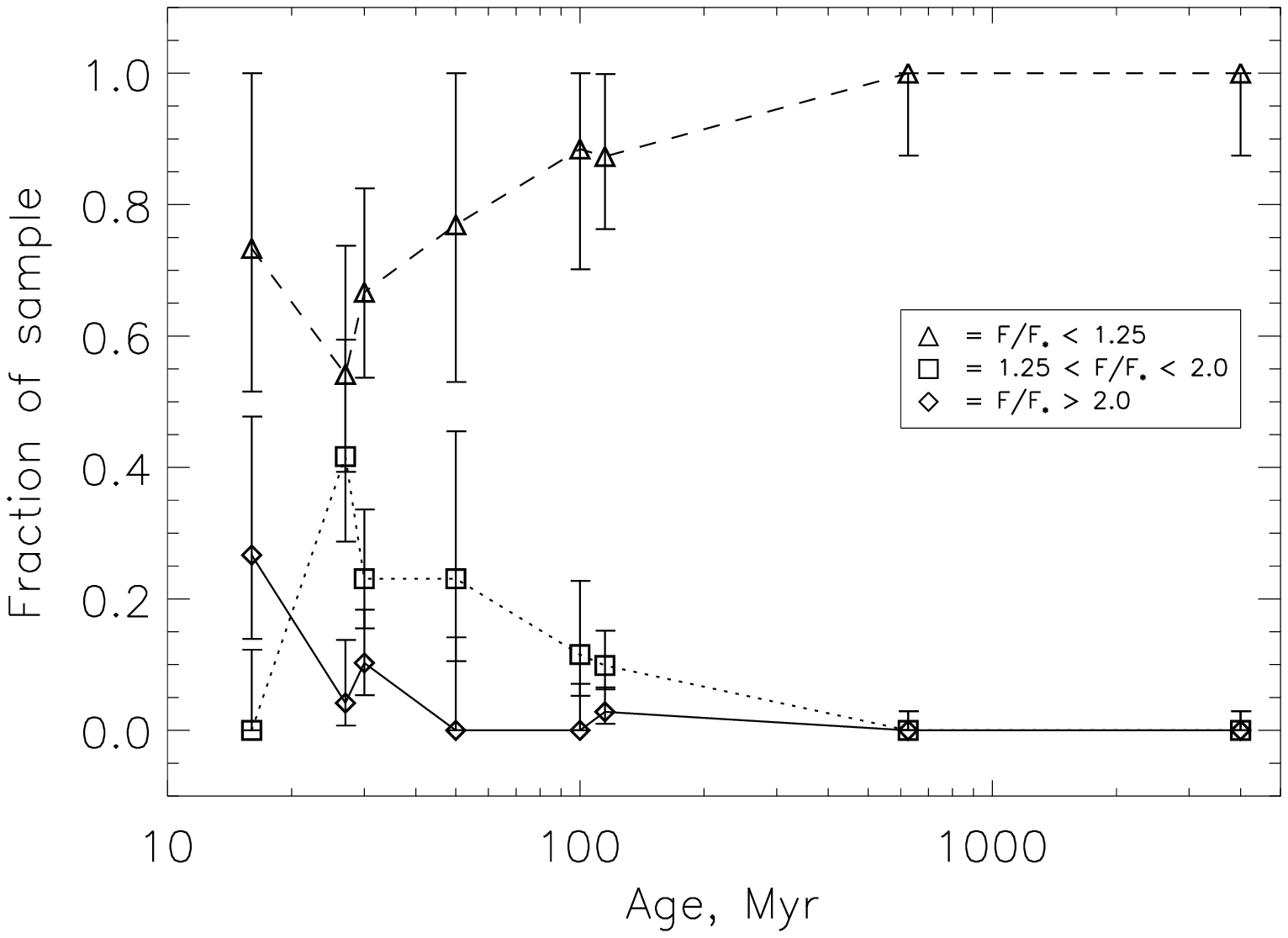}
\caption{\label{fig:xs_levels} Fraction of stars with low or no
  excess ($F_{24}/F_{\rm{phot}} < 1.25$), intermediate excess ($1.25 <
  F_{24}/F_{\rm{phot}} < 2$) and high excess ($F_{24}/F_{\rm{phot}} >
  2$) for each of the cluster samples listed in Table
  \ref{tab:cluster}.  The overall trend agrees well with the A stars
  as shown in Figure 3 of \citet{rieke} and Figure 9 of \citet{su}.
  The distribution of excess for the IC 4665 cluster presented in this
  paper (shown at 27 Myr) is somewhat different from the other cluster
  of a similar age, NGC 2547.} 
\end{figure}

We can also consider the relative proportions of different levels of
excess in the clusters studied.  We follow the example set by
\citet{rieke} who showed the rates of low, intermediate and high
excess fractions as a function of time to explore the decay of debris
discs around A stars.  We split the samples for each cluster into
small or no excess ($F_{24}/F_{\rm{phot}} < 1.25$, where
$F_{\rm{phot}}$ is the expected flux from the stellar photosphere),
intermediate excess ($1.25 < F_{24}/F_{\rm{phot}} < 2.$), and large
excess ($F_{24}/F_{\rm{phot}} > 2$). These are plotted as  
a function of age in Figure \ref{fig:xs_levels}.  The general pattern
of excess is similar to that seen around A stars, with the proportion
of targets having low or no excess ($F_{24}/F_{\rm{phot}} < 1.25$)
increasing with age, and the proportion of sources with intermediate
excess being higher than  those with large excess
($F_{24}/F_{\rm{phot}} > 2$) for all but the youngest sub-sample
(compare with Figure 3 of \citealt{rieke} and Figure 9 of
\citealt{su}). This suggests that the evolution of debris around both
solar-type stars and more massive stars follows a similar pattern,
although the timescale is an order of magnitude longer for more
massive stars (see Figure 6 of \citealt{siegler}).  

We can also see in Figure \ref{fig:xs_levels} that IC 4665 and NGC
4547 which are of the similar age appear to have somewhat different
distributions.  A higher proportion of the sources in IC 4665 have
intermediate levels of excess ($1.25 < F_{24}/F_{\rm{phot}} < 2 $)
than for any other cluster (although not significantly so).
Consequently the proportion of sources with no or low excess
($F_{24}/F_{\rm{phot}} < 1.25$) is lower than the other published
clusters (again not significantly so).  If assume that the minimum
mass solar nebula (MMSN) predictions from \citet{kb05} for the
evolution of debris in the final stages of terrestrial planet
formation can be applied to our cluster data, then the frequency of
discs with intermediate excess in IC 4665 would suggest that many
stars in the cluster have recently experienced a large
collision. However, as the initial planetesimal discs may have
differed from a MMSN then we cannot confirm this is the case. 

\begin{figure*}
\includegraphics[width=84mm]{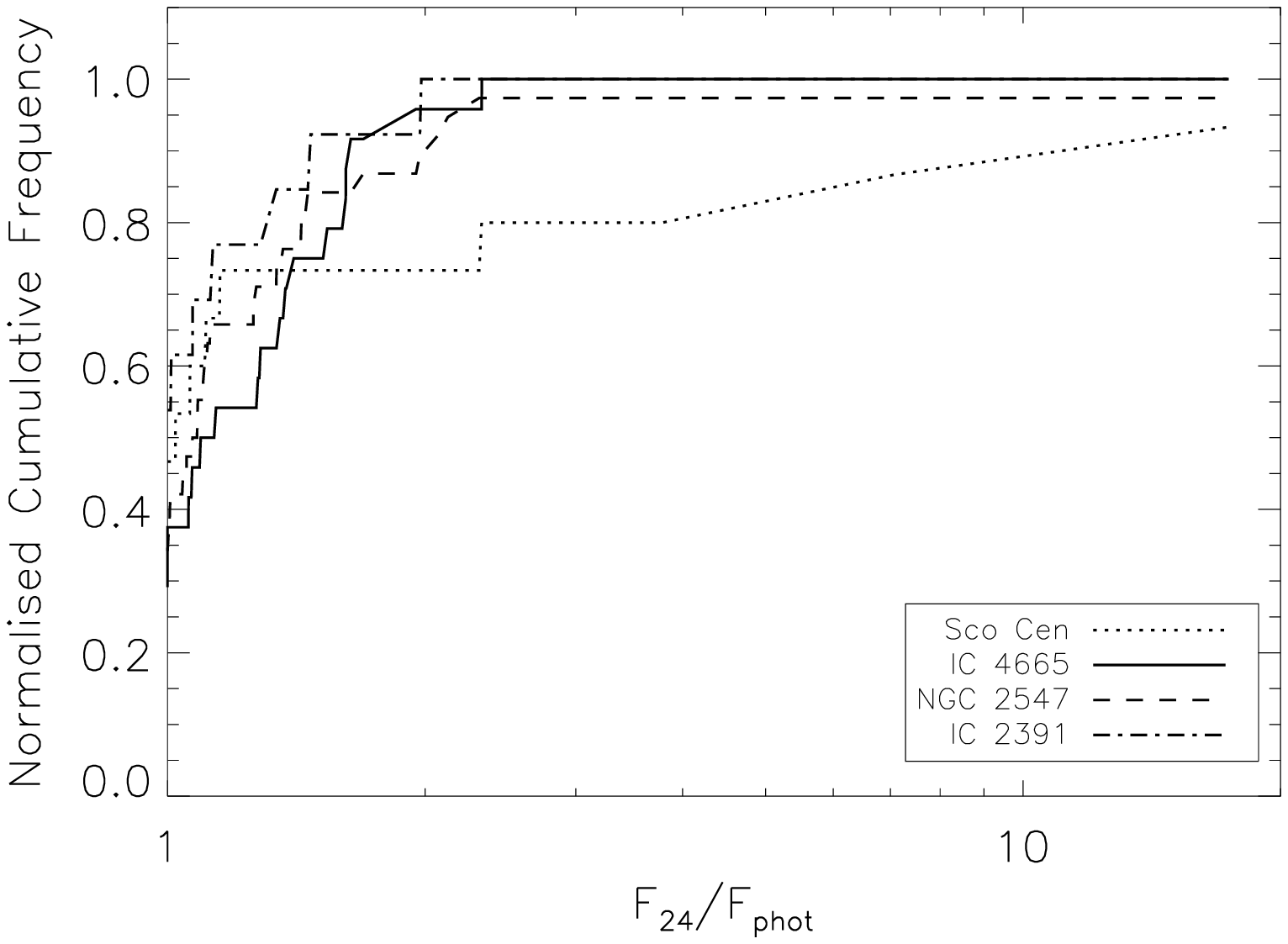}
\includegraphics[width=84mm]{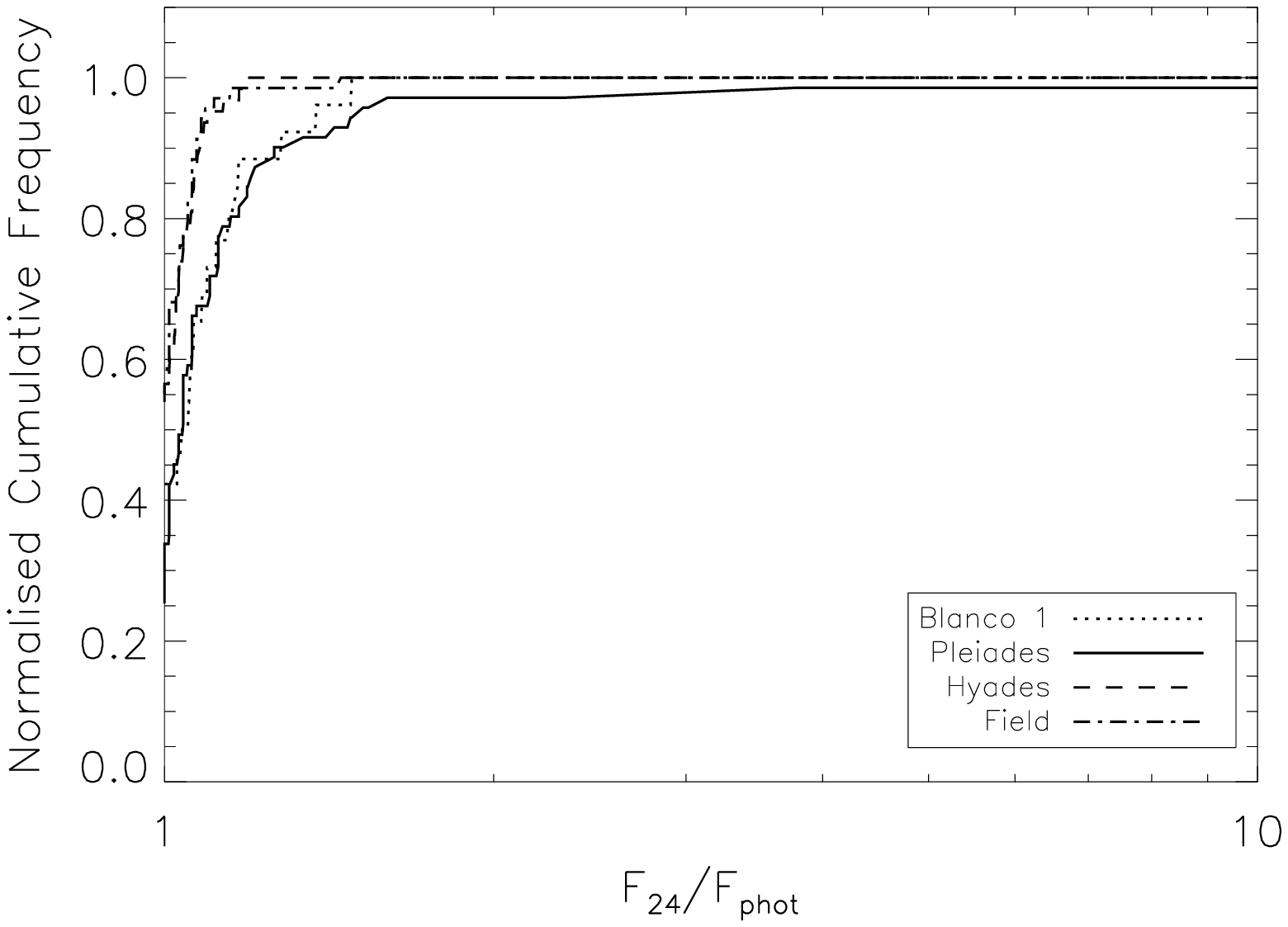}
\\
\caption{\label{fig:cum_freq} The normalised cumulative frequency
  distributions of the excess emission for the clusters listed in
  Table \ref{tab:cluster}. 
  The younger clusters are shown in the left-hand plot and the older
  clusters ($\ge$ 100Myr) in the right-hand plot.  The distribution of
  IC 4665 (solid line, left-hand plot) could be drawn from the same
  underlying distribution as the other sources $\le$ 50Myr (shown in
  the left-hand plot), but is unlikely to be drawn from the same
  distribution of excesses as the older clusters (see text for
  details). }
\end{figure*}

To further test whether the distribution of excess emission in IC 4665
is different to other sources we examine the cumulative distribution
functions of the excess emission of each cluster.  The resulting
normalised distributions are shown in Figure \ref{fig:cum_freq}.  As
we can see here the older clusters and the field have very similar
distribution functions, which are somewhat different to the younger
sources shown in the left-hand plot.  This is to be expected if the
evolutionary timescale for debris discs is of the order of 10-100Myr.
The cumulative frequency (CF) distribution for IC 4665 has some
differences to the CF distribution of NGC 2547, however a two-sample
Kolmogorov-Smirnov (K-S) test indicates that the probability that the
two cluster samples are drawn from the same underlying probability
distribution is still 76\%.  The probability that the IC 4665 and Sco Cen
samples are drawn from the same distribution is 26\%. This is not low
enough to confirm a significant difference between the clusters,
although the CF distribution looks quite different.  There is
evidence that the underlying distribution could be different for the
older samples.  A K-S test comparing IC 4665 to these samples returns
a probability that the underlying distributions are the same of 8\%
(Blanco 1, 100Myr), 1\% (Pleiades, 115Myr), 0.04\% (Hyades, 625Myr)
and 0.02\% (field sample, average age 4000Myr).  This provides further
support for evolution of debris discs on 10-100Myr timescales.

We find no evidence of a link between
binarity and excess seen in IC 4665. A binary companion limits the
size of a stable region for a circumbinary disc to $\le a_{\rm{crit}}$,
a function of the binary's configuration (e.g. \citealt{holman}). This
limit also approximately corresponds to the region in which the final
chaotic stages of planet formation from lunar-sized embryos can
proceed \citep{quintana}.  This truncation of planet formation in
binaries is of particular interest given the recent discoveries of
$\sim$ 80 exoplanets in binary systems (see e.g. \citealt{desidera},
\citealt{mugrauer}). The presence of dust grains in so-called forbidden
regions ($< a_{\rm{crit}}$) found for several binary systems by
\citet{trilling} may be explained by recent numerical
simulations in \citet{thebault} who found that small grains can
populate the forbidden region.  The amount of dust in unstable regions
depends on the balance between the rate of small grain production
through collisions and removal by the perturbations of the
binary. As discussed in section 4.3 several studies have offered
contradictory evidence for a link between multiplicity and the
presence of a debris disc (\citealt{cieza, trilling, stauffer}). With
the wealth of seemingly contradictory conclusions about the
relationship between binarity and debris disc incidence further
studies, in particular exploring the geometry of the binary system and
the true dust distributions to remove degeneracies from SED fitting,
are needed.   For example the two likely binary sources in
  IC4665 with significant excess (JC02\_373 and JC08\_257) could be
  wide binaries (several hundered AU) which would not be resolved in
  the Spitzer observations and would not be expected to affect the
  presence of 24$\mu$m excess.

\section{Conclusions}

In this paper we have presented a study of the cluster IC 4665 using
Spitzer IRAC and MIPS data.  These data have been used to search for
debris discs in the cluster.  Our conclusions are: 
\begin{itemize}
\item the cluster IC 4665 has the highest incidence of 24$\mu$m excess
  in the spectral range F5-K5 of all the clusters studied with Spitzer
  to date, although the rate ($42^{+18}_{-13}$\%) is not significantly
  higher than the similarly aged NGC 2547 ($33^{+13}_{-9}$\%).  The
  majority of the sources in the cluster have low or intermediate
  levels of excess $F_{24}/F_{\rm{phot}} < 2$. No sources in IC 4665
  have excess above the levels expected for an inverse time decay of
  debris predicted by collisional evolution models; 
\item the source TYC424-473-1, which may be a binary, has excess in
  all unsaturated observations indicative of excess emission from the
  near to mid-infrared.  Such near-infrared excess could indicate the
  presence of a remnant primordial disc. This excess can be fit by a
  simple blackbody curve at a temperature of 500K suggesting a radial
  offset of $\sim$1.7AU from the star; 
\item there is no evidence of a dependence of excess at 24$\mu$m on
  multiplicity of the star in this cluster.  We find several sources
  which are suspected multiples that have significant 24$\mu$m excess,
  but as the numbers are small these do not provide contradictory
  evidence against the recent work by \citet{stauffer} who found
  evidence that 24$\mu$m excess is reduced around multiple stars in
  other clusters.   More work to explore the nature of the multiple
  stars, in particular to determine the system geometry and
  the location and distribution of dusty debris in these 
  systems is necessary to determine the nature of any link between
  excess and multiplicity.
\end{itemize}

\section*{Acknowledgments}

\bsp

\label{lastpage}

\end{document}